\documentclass[aps, prb, superscriptaddress, preprintnumbers, floatfix, nofootinbib,longbibliography, 10pt, twocolumn]{revtex4-2}

\usepackage{graphicx, xcolor}
\usepackage{amsfonts,amsmath,amssymb,amstext,dsfont,bm}
\usepackage{float,wrapfig,subfigure, psfrag}
\usepackage{comment}
\usepackage{nicematrix}
\usepackage{amsmath}

\usepackage[colorlinks=true,urlcolor=blue,citecolor=blue,linkcolor=blue, breaklinks=true]{hyperref}
\usepackage{mathtools}

\DeclarePairedDelimiter\floor{\lfloor}{\rfloor}

\newcommand{\be}{\begin{equation}}
\newcommand{\ee}{\end{equation}}

\newcommand{\bea}{\begin{eqnarray}}
\newcommand{\eea}{\end{eqnarray}}

\renewcommand{\phi}{\varphi}
\renewcommand{\epsilon}{\varepsilon}

\graphicspath{{Figures-notes/}} 

\definecolor{purple}{rgb}{0.8,0,0.6}
\definecolor{darkgreen}{rgb}{0.00,0.6,0.00}

\definecolor{Blue}{rgb}{0,0,0.85}

\definecolor{purple}{rgb}{0.8,0,0.6}

\definecolor{orange}{rgb}{1,0.64,0}

\DeclareFontEncoding{LS1}{}{}
\DeclareFontSubstitution{LS1}{stix}{m}{n}
\DeclareSymbolFont{symbols4}{LS1}{stixbb}{m}{it}
\DeclareMathSymbol{\varhexagonblack}{\mathord}{symbols4}{"DD}
\DeclareMathSymbol{\hexagonblack}   {\mathord}{symbols4}{"DE}

\extrafloats{100}

\usepackage[flushleft]{threeparttable}
\makeatletter 
\g@addto@macro\TPT@defaults{\footnotesize} 
\makeatother
\usepackage{multirow}

\makeatletter
\let\oldfnsymbol\@fnsymbol
\def\@fnsymbol#1{\ensuremath{\ifcase#1\or \dagger\or \ddagger\or \mathsection\or \mathparagraph\or \|\or \dagger\dagger\or \ddagger\ddagger\else\@ctrerr\fi}}
\makeatother

\allowdisplaybreaks

\begin{document}

\title{Electronic properties and topological aspects of graphene nanohelicoids}

    \author{Xiaoqian Liu\textsuperscript{*}}
    \email{xiaoqian.liu@epfl.ch}
    \affiliation{Institute of Physics, \'{E}cole Polytechnique F\'{e}d\'{e}rale de Lausanne (EPFL), CH-1015 Lausanne, Switzerland}

    \author{Arsen Herasymchuk\textsuperscript{*}}
    \email{arsen.herasymchuk@uzh.ch}
    \affiliation{Department of Physics, University of Z\"{u}rich,  CH-8057 Z\"{u}rich, Switzerland}

    \author{Yaroslav Zhumagulov}
	\affiliation{Institute of Physics, \'{E}cole Polytechnique F\'{e}d\'{e}rale de Lausanne (EPFL), CH-1015 Lausanne, Switzerland}
	
	\author{Oleg V. Yazyev}
	\email{oleg.yazyev@epfl.ch}
	\affiliation{Institute of Physics, \'{E}cole Polytechnique F\'{e}d\'{e}rale de Lausanne (EPFL), CH-1015 Lausanne, Switzerland}

\date{\today}
 
\begin{abstract}

We introduce graphene nanohelicoids, geometric analogs of graphene nanoribbons, in which the honeycomb lattice is embedded on a helicoidal surface. Starting from the three-dimensional helical structure, we construct effective one-dimensional lattice models with band structures characterized by a momentum-shifted particle–hole relation $E_v(k)=-E_c(k+\pi)$ that reflects an antichiral symmetry arising from the nonsymmorphic symmetry. A systematic investigation of graphene nanohelicoids using the tight-binding approximation reveals a number of trends upon varying width and edge orientation, for instance, alternating transitions between semiconducting and metallic regimes. As the structure width varies, the band gap periodically closes and reopens, accompanied by an alternating Zak phase that switches between trivial and nontrivial. Using an analytic tight-binding model with a continuous deformation of the graphene nanohelicoids, we explain the origin of width- and termination-dependent band inversion and alternating Zak phase, and derive the relationship between the termination-resolved index and the sublattice symmetry operator in one-dimensional nonsymmorphic systems.

\end{abstract}

\maketitle

\makeatletter
\let\@fnsymbol\oldfnsymbol
\makeatother

\renewcommand{\thefootnote}{\fnsymbol{footnote}}
\footnotetext{These authors contributed equally.}
\renewcommand{\thefootnote}{\arabic{footnote}}
\setcounter{footnote}{0}

\section{Introduction}

Graphene nanoribbons have made a long way from starting as a highly popular theoretical model \cite{Fujita96,Nakada96,Son06,Barone06} to becoming a versatile experimental platform for controlling the electronic structure of graphene through geometry and boundary conditions~\cite{Cai10,Tao11,wang-nanoribbon-natrevphys21,Zhang26}. 
Unlike infinite two-dimensional graphene, a graphene nanoribbon has a finite width and well-defined edges, so that its low-energy spectrum is determined not only by the bulk Dirac bands, but also by boundary conditions and lattice termination~\cite{Klein99,Ryu02,wassmann-ribbons-prl08,cao-z2-prl17}.
Graphene nanoribbons with zigzag or low-symmetry edges support edge-localized states near the Fermi level, whereas armchair nanoribbons generally do not host such robust edge states. Their low-energy gaps are instead determined by how the finite ribbon width fits with the underlying graphene lattice~\cite{Nakada96, Son06,cao-z2-prl17}.
Thus, graphene nanoribbons are not merely finite-size fragments of graphene, but effective one-dimensional (1D) systems whose low-energy spectra are shaped by their finite width~\cite{Nakada96,Son06}, crystallographic orientation of the edges~\cite{Akhmerov08,delplace-zakphase-prb11,Mong11,Yazyev11}, and the details of their atomic structure~\cite{wassmann-ribbons-prl08,rizzo-nanoribbon-nature18,mangnus-topological-gnr-prb22,sakaguchi-gnr-catalyst-natcom24} (for review, see Ref.~\onlinecite{Yazyev13}).
The same sensitivity to width, edge geometry, and registry also controls transport, contact, and junction physics in nanoribbon devices~\cite{nakabayashi-band-filter-prl09,verges-armchair-contact-prb18,rizzo-graphene-science20,Ceernevics20,tran-zigzag-parity-prb24,Cernevics24,Leuenberger26}.
In addition, edge magnetism, charge doping, interaction effects, and chemically engineered edge motifs provide further routes to modify nanoribbon states~\cite{Okada01,Son06b,Pisani07,Yazyev08,Tao11,zhang-chiral-magnetic-prb17,slota-magnetic-edge-nature18,slota-magnetic-edge-nature18,luo-zigzag-magnetism-prb20,michele-nanoribbon-jpcl25,Song25}.

The same idea of geometry-controlled electronic structure also appears in graphitic systems with extended lattice defects, which provide a route to engineer electronic properties through geometry rather than composition~\cite{oleg-defect-prb10,Yakobson10,banhart-defect-acsnano11,butz-dislocation-nature14,dai-dislocation-prb16,lin-defect-natrev23}. 
In graphene systems, topological defects such as dislocations~\cite{oleg-defect-prb10,Yakobson10,butz-dislocation-nature14,dai-dislocation-prb16}, grain boundaries~\cite{oleg-defect-prb10,Yakobson10,malola-grain-boundary-prb10,banhart-defect-acsnano11}, and strain solitons~\cite{alden-strain-pnas13,bultinck-tbg-strain-prl21} have long been recognized as important sources of structural and electronic reconstruction~\cite{neto-graphene-rmp09,banhart-defect-acsnano11,sergey-nm14}.
Among these, screw dislocations represent a particularly interesting class of defects in multilayer graphene and graphite, in which the stacking of graphene sheets is continuously shifted along a crystallographic direction~\cite{hennig-screw-science65,patel-screw-bjap65,chen-screw-acsnano14,shearer-screw-jacs17,zhao-screw-amr22}.
In three-dimensional crystals, a screw dislocation is characterized by a Burgers vector parallel to the dislocation line, in contrast to an edge dislocation, where the Burgers vector is perpendicular to the dislocation line~\cite{slager-topo-prb14,zhao-screw-amr22,lin-defect-natrev23}.
Screw dislocations in graphite were identified already in early studies~\cite{hennig-screw-science65,patel-screw-bjap65}, and have since been widely investigated as a mechanism for generating nontrivial layer stacking~\cite{chen-screw-acsnano14,shearer-screw-jacs17}, spiral growth~\cite{tay-spiralgrowth-cm18,zhao-screw-nl21,zhao-screw-amr22,wang-spiral-natmat24}, and moir\'e structures~\cite{liu-spiral-prl24,wang-spiral-natcom24}.
More broadly, screw dislocations have emerged as routes to topological responses across electronic and photonic systems~\cite{lustig-dislocation-nature22,zhou-screw-prb25}, acoustic analogs~\cite{ye-3dacoustic-natcom22}, and phononic systems~\cite{wang-phonon-screw-prb21,zhou-phonon-screw-afm25}, where geometry can modify transport, band connectivity, and boundary localization.

From the perspective of graphene nanoribbon physics, these developments naturally suggest a geometric extension of nanoribbons from a flat strip to a helicoidal structure, connecting curved or compactified nanoribbon geometries~\cite{gueclue-mobius-ribbon-prb13,gong-compacted-dimensions-prb23} with graphene-helicoid models~\cite{PhysRevB.92.205425,PhysRevB.92.035440,zhan-helicoid-carbon17,zhan-helicoid-jpcc18,balakrishnan-localization-helical-prb25}.
This direction is related to spiral graphite and screw-dislocation structures~\cite{xu-spiral-nl16,tay-spiralgrowth-cm18,zhao-screw-amr22}, including recent spiral-graphite and moir\'e realizations~\cite{wang-spiral-natmat24,wang-spiral-natcom24,liu-spiral-prl24}. However, the system considered here is not spiral graphite in the conventional sense, for which a layered system takes the finite interlayer interaction into account.
Instead, we take the graphene out of the graphite: The lattice is embedded in a helicoidal geometry, and only the intralayer coupling is considered; the resulting system should be viewed as the graphene helicoid~\cite{PhysRevB.92.205425,PhysRevB.92.035440,zhan-helicoid-carbon17,zhan-helicoid-jpcc18}, related more broadly to Riemann-surface and helical structures~\cite{xu-spiral-nl16,liu-helical-nature19,sergey-helix-arxiv25}. 
Finite-length graphene helicoidal structures are represented by helicenes, a family of polycyclic aromatic molecules with unique optical properties stemming from their structural chirality~\cite{Helicene1,Helicene2,Helicene3,Helicene4,Helicene5}.

This distinction is particularly important for monolayer graphene~\cite{neto-graphene-rmp09}. 
For strictly two-dimensional, flat graphene, conventional screw dislocations are not naturally defined because there is no out-of-plane dislocation line with a parallel Burgers vector~\cite{oleg-defect-prb10,zhao-screw-amr22}. A helicoidal embedding instead imposes a screw-type global boundary condition while preserving the local graphene connectivity: Motion around the circumference is tied to translation along the helical axis, effectively placing the electronic problem on a helicoid manifold~\cite{PhysRevB.92.205425,PhysRevB.92.035440,xu-spiral-nl16}.
For a finite-width helicoid, the circumference defines a characteristic transverse size, so the low-energy physics acquires a quasi-one-dimensional character and the zigzag or armchair edge termination remains decisive, as in graphene nanoribbons~\cite{Tao11,akbari-zigzag-width-prb16}. At the same time, the bipartite lattice and screw translation symmetry introduce unconventional nonsymmorphic constraints~\cite{nonsymm-shengyuan-prb18,zhang-nonsymmorphic-nanoribbon-prb18,nonsymm-zhang-prl23}, which can influence band connectivity, boundary states, and defect-localized responses~\cite{cao-z2-prl17,zhou-screw-prb25}.
In this sense, graphene nanohelicoids (GNHs) can be viewed as the helicoidal counterparts of graphene nanoribbons. Locally, they still keep the honeycomb lattice, finite width, and well-defined zigzag or armchair edge of graphene nanoribbon. Globally, they are embedded in helical surfaces, linking rotational motion and translational motion along the helical axis~\cite{PhysRevB.92.205425,PhysRevB.92.035440,zhan-helicoid-carbon17,zhan-helicoid-jpcc18,liu-helical-nature19}.
Therefore, GNHs combine the nanoribbon boundary physics with the screw symmetry, providing a concise platform for investigating geometry-driven electronic and topological properties.

In our work, we primarily discuss the electronic properties and topology of graphene nanohelicoids, providing an effective one-dimensional description. Starting from the three-dimensional helicoidal geometry, we construct a one-dimensional lattice model and show that the resulting nearest-neighbor band structure exhibits an antichiral symmetry, i.e., a momentum-shifted spectral relation that constrains band connectivity in a manner reminiscent of nonsymmorphic systems~\cite{nonsymm-zhang-prl23}. Importantly, this symmetry is rooted in the antibipartite lattice structure of the effective model. We then investigate how boundary conditions enter this picture, where the interplay between helical quantization and lattice symmetry can be analyzed in a controlled and transparent way.

\begin{figure}[t]
	\centering
	\includegraphics[width=8.6cm]{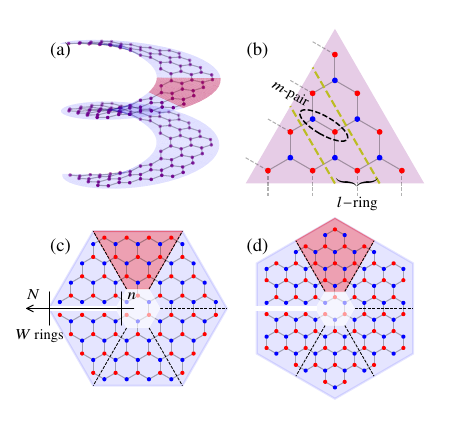}
	\caption{(a) Graphene nanohelicoid lattice model. (b) Unit cell of zigzag-edge helicoid. Red/blue dots denote the $A$/$B$ sublattice. Grey dashed lines show the connectivity of sites with neighboring unit cells. 
    (c) Lattice model of the zigzag-edge 1D helicoid lattice model. $N$ is the outer radius and $n$ is the inner radius, $W=N-n+1$ is the width of the helicoid flake. (d) Armchair-edge helicoid lattice model. 
}\label{fig:spiral_lattice}
\end{figure}

We further demonstrate that the discretized finite width serves as a direct geometric knob for tuning the low-energy spectrum, analogous to the width-dependent electronic structure of graphene nanoribbons. As the sample width varies, the system undergoes repeated transitions between semiconducting and metallic regimes through symmetry-constrained gap closings and openings. These transitions are accompanied by an alternation of the Zak phase between trivial and nontrivial values, revealing a width-controlled switching of bulk polarization in the effective one-dimensional bands~\cite{zakphase-prl89,delplace-zakphase-prb11}. To elucidate the mechanism, we first derive an analytic solution for the winding number of the GNH Hamiltonian within a tight-binding model, showing the emergence of an alternating Zak phase. We then introduce an adiabatic deformation of graphene nanohelicoids, which allows a demonstration of the alternating Zak phase in the atomic limit.

This paper is organized as follows. In Sec.~\ref{sec:band}, we introduce the lattice structure of the graphene helicoid, and derive its one-dimensional tight-binding description. We establish the anti-bipartite model and discuss its band structure, identifying the dependence of the energy gap on system width, with an emphasis on zigzag boundary conditions. 
In Sec.~\ref{sec:ldos}, we present the numerical results for the local density of states associated with the band structure of the graphene helicoid, considering both zigzag and armchair boundary conditions. In Sec.~\ref{sec:zak}, we provide an analytic solution of the winding number and the Zak phase for the tight-binding model and analyze its dependence on the termination and the helicoid geometry. The results are summarized in Sec.~\ref{sec:discussion}. Technical details are presented in Appendixes~\ref{sec:schur-proof} and \ref{sec:real-proof}.

\section{Antichiral symmetry and band structure} \label{sec:band}

Let us consider the symmetry of the lattice structure of the GNH on the example of the zigzag edge. Its schematic lattice model is presented in Fig.~\ref{fig:spiral_lattice}. The GNH exhibits a fractional translation symmetry that combines a rotation in the $xy$-plane with a fractional translation along the $z$-axis, as shown in Fig.~\ref{fig:spiral_lattice}(a). Therefore, the nonsymmorphic symmetry of such a system can be denoted as $\left\{C_{6} | \frac{c}{6}\right\}$, where $c$ is the full unit cell lattice constant in the $z$-direction~\cite{nonsymm-zhang-prl23}. The width of the helicoid is given by $W=N-n+1$, where $N$ and $n$ denote the outer and inner radii, respectively. 
Assuming the equal couplings between the atoms, the unit cell can be therefore constructed as the one triangular sector, which is highlighted in red in Fig.~\ref{fig:spiral_lattice}(c), which encodes the nonsymmophic symmetry of the initial system in its effective 1D description.

For such a convention of the unit cell, the notion of antibipartite symmetry is introduced. The unit cell possesses the chiral symmetry with non-equal sublattices~\cite{dai2024topologicalclassificationchiralsymmetry}, where the chiral symmetry itself can be seen explicitly from the band structure if the interchange of the sublattices is included in the translation operator.
As a result for the regular translation operator convention, the nearest neighbor band structure obeys a momentum-offset particle-hole relation rather than a conventional onsite chiral constraint. In contrast to standard chiral models, where chiral symmetry enforces the valence band and conduction band $E_v(k)=-E_c(-k)$ and leads to symmetric band structures about zero energy, antichiral symmetry (which is used as a notion for the chiral symmetry for anti-bipartite systems) implies a spectral relation of the form $E_v(k)=-E_c(-k+Q)$, with $Q$ a fixed reciprocal-lattice vector. Such symmetry typically arises in lattices with nonsymmorphic crystalline operations that combine sublattice exchange with fractional translation~\cite{nonsymm-shengyuan-prb18, nonsymm-zhang-prl23}. If in addition the time-reversal symmetry is preserved, the latter spectral relation is $E_v(k)=-E_c(k+Q)$.

\begin{figure}[t]
	\centering
	\includegraphics[width=7.2cm]{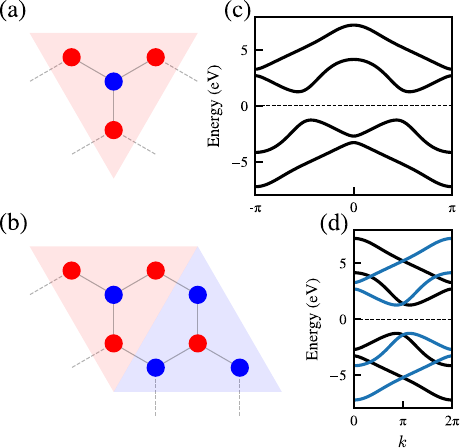}
	\caption{Comparison between bipartite and anti-bipartite lattice models and their corresponding band structure; $N=2$ and $n=1$ case. (a) The antibipartite lattice composed of two sublattices (red and blue). (b) The bipartite lattice model with the double unit cell. (c) The antichiral band structure for the corresponding lattice. (d) The chiral band structure with folded Brillouin zone. The blue curves highlight pairs of bands related by this momentum-shifted particle–hole constraint.}\label{fig:bipartite}
\end{figure}
    
As shown for the $W=2$ ($N=2$, $n=1$) case in Fig.~\ref{fig:bipartite}(a), the unit cell of the GNH for the anti-bipartite lattice contains three $A$ sublattice sites (red) and one $B$ sublattice site (blue). This imbalance represents a chiral symmetry with non-equal sublattices~\cite{dai2024topologicalclassificationchiralsymmetry}. 
Consequently, the band structure does not satisfy the standard relation $E_v(k)=-E_c(k)$. Instead, as illustrated in Fig.~\ref{fig:bipartite}(c), the spectrum obeys a momentum-shifted particle–hole symmetry of the form $E_v(k)=-E_c(k+\pi)$, characteristic of an antichiral lattice. In contrast, the chiral lattice model shown in Fig.~\ref{fig:bipartite}(b) preserves conventional bipartite symmetry, leading to the spectral constraint $E_v(k)=-E_c(k)$ [Fig.~\ref{fig:bipartite}(d)]. Notably, this chiral lattice can be viewed as a doubled supercell of the antichiral model. Correspondingly, its band structure is obtained by Brillouin-zone folding of the antichiral spectrum, i.e., the chiral bands arise from downfolding the momentum-shifted band pairs of the antichiral lattice.

A notable feature is that the band gap closure or opening is highly sensitive to the number of sites in the primitive cell. When the number of atoms is even, the spectrum generally exhibits a sizable gap around the Fermi level. In contrast, when the number of atoms is odd, the band structure becomes gapless, and a band necessarily crosses the Fermi level. 
The gap closing and reopening can be understood in terms of an effective pseudopairing structure imposed by the helical geometry. In ordinary graphene, the $A$ and $B$ sublattices are well-defined, and the relative phase between their Bloch-wave amplitudes forms the pseudospin texture responsible for Dirac physics. In the GNH, however, a global $A$/$B$ sublattice assignment is no longer exact because of the anti-bipartite connectivity. Instead, the spectrum exhibits a momentum-shifted particle-hole symmetry, indicating that states are paired between momenta separated by $\pi$. For configurations with an even number of atoms, all states can be pseudopaired, allowing a gap to open. For odd configurations, one state remains unpaired, leading to an isolated band to cross the Fermi level.

\begin{figure}[t]
	\centering
	\includegraphics[width=8.6cm]{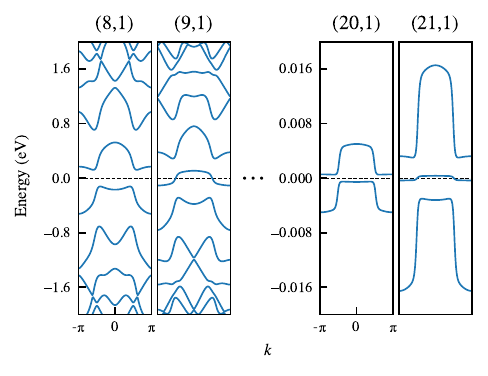}
	\caption{Band structures of zigzag-edge GNHs for different system sizes as a function of the dimensionless Bloch momentum $k$ along the periodic direction. The numbers shown above each panel denote the pair $(N, n)$. In this figure, the inner radius $n$ is kept fixed while the outer radius $N$ is varied. 
    }\label{fig:zigzag_band}
\end{figure}

For the antibipartite lattice with zigzag edge, we construct the unit cell as demonstrated in Fig.~\ref{fig:spiral_lattice}(b). The atoms are labeled in the way $\left\{ A_{m,l},B_{m,l}\right\}$, describing $m$-pairs of $A,B$ atoms on the $l$-ring of the helicoid. Note that, in the antibipartite formulation, the atoms $\left\{ A_{l,l}\right\}$ do not have a corresponding pair with $\left\{ B_{l,l}\right\}$.
Therefore, the tight-binding Hamiltonian of this system on $l^2(\mathbb{Z})$ reads as follows:
\begin{equation}
\label{eq:hamiltonian-def-anti}
\begin{aligned}
\hat{H}&=  t \sum_{i \in \mathbb{Z}} \Bigg\{ \sum_{l=n}^{N}\sum_{m=1}^{l-1}  \Big[ \hat{a}^{\dagger}_{i,m,l}\hat{b}_{i,m,l} \\
&+  \hat{b}^{\dagger}_{i,m,l}\hat{a}_{i,m+1,l} \Big]+\sum_{l=n}^{N}   \hat{a}^{\dagger}_{i,l,l}\hat{a}_{i+1,1,l}   \\
&+ \sum_{l=n}^{N-1}\sum_{\left<  m_1,m_2 \right>}   \hat{a}^{\dagger}_{i,m_1,l}\hat{b}_{i,m_2,l+1}  +\text{h.c.} \Bigg\} , 
\end{aligned}
\end{equation}  
where the summation over $\left< m_1 ,m_2 \right>$ is between the nearest $A$ atom of $l$-ring and $B$ atom of $(l+1)$-ring. Here, $t$ is the intralayer hopping parameter.
The band structure is presented in Fig.~\ref{fig:zigzag_band}. For the numerical calculation, we set the hopping parameter $t=2.7~\mbox{eV}$ with the in-plane lattice constant in graphene $a=2.46~\mbox{\AA}$~\cite{tbmodel-sk-method}. The unit cell is constructed in a way to map the nonsymmorphic symmetry of the structure, presented in Fig.~\ref{fig:spiral_lattice}(a), into the translational symmetry of the effective one dimensional chain. Here, $k\in\left[-\pi,\pi\right]$ is the dimensionless phase, obtained from the Fourier transformation over $i$ in Eq.~(\ref{eq:hamiltonian-def-anti}), and should be distinguished from the usual notion of quasiparticle pseudomomentum.

We first focus on the zigzag edge configuration, which provides the simplest setting to analyze the band structure of the model. In graphene-based systems, zigzag and armchair edges lead to qualitatively different electronic spectra because they impose different boundary conditions on the lattice. In particular, zigzag edges terminate one sublattice at the boundary and therefore preserve the valley degree of freedom, while the armchair edges couple the two valleys $K$ and $K'$ through the boundary condition.

As shown in Fig.~\ref{fig:zigzag_band}, the band gap depends on the parity of the number of rings $W$ in the zigzag configuration. In this case, the primitive cell contains $W\left( W+2 n - 2 \right)$ atoms. Consequently, systems with even $W$ have an even number of sites in the effective one-dimensional unit cell and exhibit a finite gap near the Fermi level, whereas systems with odd $W$ necessarily host a band crossing and remain gapless. The gapless band of odd $W$ graphene helidcoids might resemble chiral anomalies; however, it corresponds to the different mechanisms. For example, in Janus GNRs, the chiral modes are generated from electric field induced sublattice-imbalance~\cite{saroka-chiral-anomaly-prb24, saroka-chiral-anomaly-nanolett26}. However, in our case, the band crossing and sublattice imbalance come from the odd number of sites in the screw-primitive unit cell itself rather than from the external field. This even–odd behavior is a direct consequence of the antibipartite lattice structure discussed previously.

More importantly, the band dispersion evolves systematically with increasing helicoid width. As illustrated in Fig.~\ref{fig:zigzag_band}, the band gap decreases as $W$ increases and the bands gradually become flatter. In the large-width limit, the spectrum approaches a step-like profile, where the energy remains nearly constant over a wide range of momenta and varies rapidly only near the band edges. A useful way to understand this behavior is to relate the one-dimensional bands to the Dirac-like dispersion of graphene. Near the Dirac points, the graphene spectrum is approximately linear, $E(\mathbf{k}) \approx v_F |\mathbf{k}|$. When this cone-like dispersion is examined along a single momentum direction, the resulting energy varies weakly over most of the Brillouin zone and changes more rapidly near the band edges. As the helicoid width increases, the effective dispersion along the helicoid axis becomes progressively flatter, giving rise to the step-like spectral profile observed in Fig.~\ref{fig:zigzag_band}.

\begin{figure}[t]
	\centering
	\includegraphics[width=8.6cm]{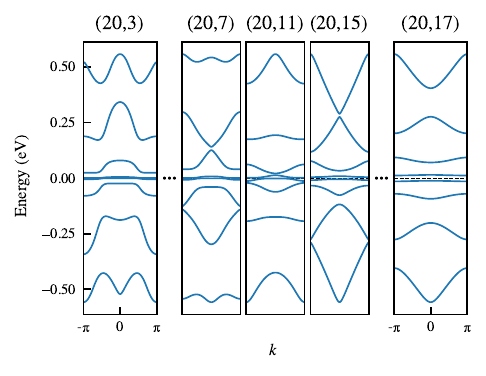}
	\caption{Band structures of zigzag-edge GNHs for different system sizes. In this figure, the outer radius $N$ is kept fixed while the inner radius $n$ is varied.
    }\label{fig:zigzag_vacuum}
\end{figure}

We next consider the case where the outer radius is kept fixed and the inner radius is varied, as illustrated in Fig.~\ref{fig:zigzag_vacuum}. In this geometry, a second zigzag edge appears at the inner edge of the structure. As the inner radius increases, the band structure near the Fermi level changes qualitatively: The step-like dispersion observed in the absence of the central hole in the helicoid gradually disappears, the bands become nearly flat, and the band gap increases.
This behavior can be understood from the increasing dominance of the zigzag edge states in the low-energy spectrum. When the central region is removed, the effective bulk area connecting the inner and outer edges is reduced, and the electronic states near the Fermi level become strongly confined to the boundaries. In the nearest-neighbor tight-binding model, the zigzag edge states are highly localized and largely sublattice polarized, which intrinsically exhibit very weak dispersion. As a result, once the spectrum becomes dominated by these edge states, the bands appear nearly flat, and the gap increases with inner radius. Consequently, the resulting band structure differs from the conventional graphene nanoribbons or nanotubes, where translational periodicity along the ribbon axis allows electronic states to propagate as extended Bloch waves, producing strongly dispersive bands.

\begin{figure}[t]
	\centering
	\includegraphics[width=8.6cm]{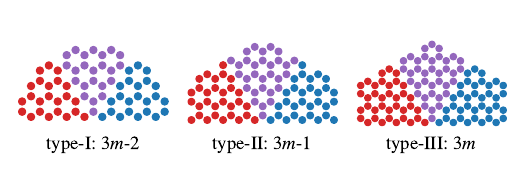}
	\caption{Schematic lattice structure of different types of armchair-edge GNHs with the corresponding index. Here, $m\in \mathbb{N}$. 
    }\label{fig:armchair_type}
\end{figure}

In contrast to the zigzag case, the band structures of armchair GNHs are less sensitive to the system size. However, the armchair geometry is not characterized by a single continuous family of terminations.
Instead, depending on the discrete index, three inequivalent armchair-zigzag junctions within a unit cell appear. Here, the index in the armchair-edge GNH serves as a width counterpart of the zigzag-edge GNH, therefore its precise local registry depends on the discrete armchair width. In the large width situation, the armchair helicoid is globally armchair terminated, but its low-energy spectrum can still be affected by the local zigzag junction.
\begin{figure}[t]
	\centering
	\includegraphics[width=8.6cm]{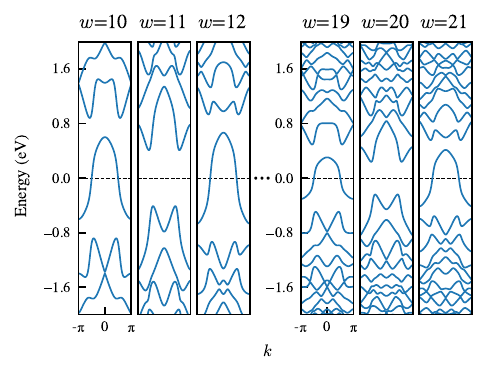}
	\caption{Band structures of armchair-edge GNHs for different system sizes. The number shown above each panel denotes the index of the armchair configuration [see Eq.~(\ref{eq:armchair-index})].
    }\label{fig:armchair_band}
\end{figure}
For concreteness, we would consider the case without the inner hole for the armchair case in this section.
For the armchair edge case, one can see three different types of junctions between the unit cells as presented in Fig.~\ref{fig:armchair_type}. Such configurations are realized for the following indices of the armchair structure:
\begin{equation}
\label{eq:armchair-index}
\begin{aligned}
w^{I}&= 3m - 2 ,\\
w^{II}&= 3m -1 ,\\
w^{III}&= 3m  ,
\end{aligned}
\end{equation}
with $m\in \mathbb{N}$. The index defines the number of armchair-type rings in the helicoid structure, analogous to width in the zigzag case. The corresponding number of atoms in the primitive cell for each type is given by:
\begin{equation}
\begin{aligned}
N^{I}&= 3m^2-m-1 ,\\
N^{II}&= 3m^2 +m  ,\\
N^{III}&= 3m^2+3m+1 .
\end{aligned}
\end{equation}
Hence, systems with index $w=w^{II}$ have an even number of sites in the effective one-dimensional unit cell and exhibit a finite gap near the Fermi level, whereas systems with index $w=w^{I}$ or $w=w^{III}$ necessarily host a band crossing and remain gapless, which is demonstrated in Fig.~\ref{fig:armchair_band}. 

The low-energy spectrum of armchair GNHs is controlled primarily by the discrete junction type rather than by the continuous increase of the helicoid width. 
This is qualitatively different from the zigzag case, where edge-localized states strongly affect the dispersion and lead to a pronounced width dependence. 
This behavior can be attributed to the boundary conditions imposed by the armchair termination. 
Unlike zigzag boundaries, armchair boundaries do not support robust edge-localized zero-energy modes. 
Consequently, changing the helicoid width mainly increases the number of bulk-like subbands without qualitatively modifying the low-energy dispersion. 
The dominant qualitative distinction is therefore the discrete armchair junction class: Type-II structures are gapped, while type-I and type-III structures remain gapless because of the odd number of sites in the effective one-dimensional unit cell.

\begin{figure}[t]
	\centering
	\includegraphics[width=8.0cm]{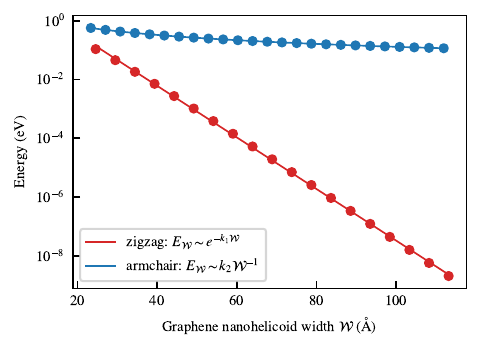}
	\caption{Band gap as a function of GNH real space width $\mathcal{W}$, where $\mathcal{W}=Wa$ for zigzag case and $\mathcal{W}=(w^{II}/2+1)a$ for armchair case. The dots are the original data, and the lines are the fitting curves.
    }\label{fig:bandwidth}
\end{figure}

For the zigzag edge, the band gap decreases rapidly with increasing helicoid width, as shown in Fig.~\ref{fig:bandwidth}.
The red dots are the zigzag edge band gap original data, and the line is the fitting curve. The log scaling fitting line indicates an approximately exponential dependence of the form $E_\mathcal{W}\propto e^{-k_1 \mathcal{W}}$, $k_1>0$.
This behavior originates from the edge-localized states associated with zigzag terminations. In graphene systems, zigzag edges host modes whose wavefunctions decay exponentially away from the edge~\cite{saroka-nanoribbon-pss14, arnold-nanoribbon-prl22}, so that the dispersion of the corresponding one-dimensional bands arises from the hybridization of these edge states along the helicoid direction. Since the effective coupling between such states is determined by the overlap of exponentially decaying wavefunctions, the resulting band gap decreases exponentially as the helicoid width increases.

The situation is qualitatively different for the armchair edge. In this case, the band gap follows a power-law dependence $E_\mathcal{W}\propto k_2\mathcal{W}^{-1}$, $k_2 >0$, as shown in Fig.~\ref{fig:bandwidth}. Unlike zigzag edges, armchair edges do not support localized edge modes; instead, the low-energy states originate from confined bulk Dirac states. The armchair edge mixes the $K$ and $K'$ valleys and imposes standing-wave boundary conditions, resulting in finite-size quantization of the Dirac spectrum. Since the graphene dispersion near the Dirac point is linear, and the quantized momentum scales as $k\sim 1/\mathcal{W}$, the characteristic energy scale of the confined states follows $E\sim \hbar v_F/\mathcal{W}$. Consequently, the band gap decreases approximately as $1/\mathcal{W}$, which is characteristic of Dirac quantum-dot–like confinement in finite graphene structures.

\section{Local density of states} \label{sec:ldos}

\begin{figure}[t]
	\centering
	\includegraphics[width=8.6cm]{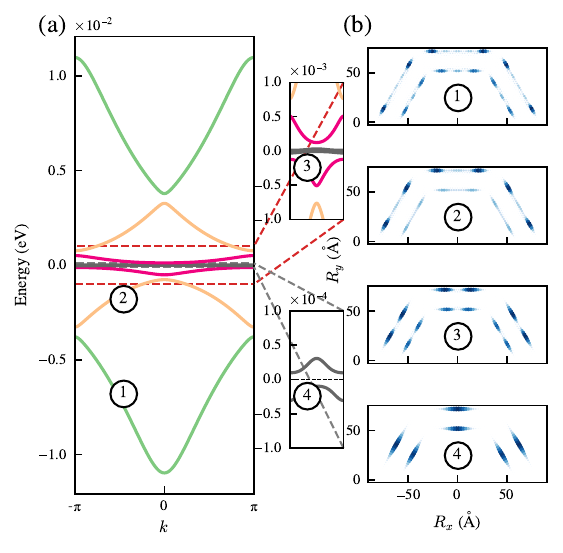}
	\caption{Band structure and corresponding LDOS for a zigzag-edge graphene helicoid with outer radius $N=34$ and inner radius $n=25$.
    (a) The band structure, with two inset panels highlighting the bands closest to the Fermi level. (b) The LDOS corresponding to the bands labeled 1–4 in the band structure.}\label{fig:zigzag_ldos}
\end{figure}

\begin{figure}[t]
	\centering
	\includegraphics[width=8.6cm]{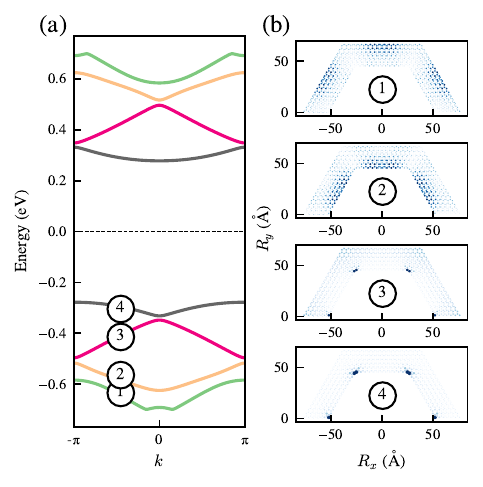}
	\caption{Band structure and corresponding LDOS for a gapped armchair outer and inner edge flake with outer index $w_{\text{out}}=54$ and inner index $w_{\text{in}}=37$.
    The outer edge corresponds to the type-III armchair edge while the inner edge corresponds to the type-I armchair edge (see Fig.~\ref{fig:armchair_type}). (a) The band structure with indices. (b) The LDOS corresponding to the bands labeled 1–4 in the band structure.    }
    \label{fig:armchair_ldos}
\end{figure}

To further investigate the origin of the low-energy spectrum, we analyze the local density of states (LDOS) associated with the band structure shown in Fig.~\ref{fig:zigzag_ldos}. The system considered here has outer radius $N=34$ and inner radius $n=25$. The two inset panels highlight the bands closest to the Fermi level, while Fig.~\ref{fig:zigzag_ldos}(b) displays the corresponding LDOS distributions labeled by the band indices. A clear correspondence between the band structure and spatial localization can be observed. The LDOS associated with the bands near the Fermi level is strongly concentrated along the zigzag edges of the helicoid flake.
In contrast, the LDOS of bands farther away from the Fermi energy spreads over the interior region of the lattice, revealing their bulk-like character.

These results demonstrate that the low-energy electronic structure of the zigzag helicoid system is primarily governed by zigzag edge states, whereas higher-energy bands originate from extended bulk states. Because the edge states are spatially confined along the boundaries, their dispersion is significantly reduced compared with the bulk-derived bands, consistent with the nearly flat bands discussed in the previous section. This clear separation between edge-localized modes and bulk states may provide a natural starting point for constructing an effective low-energy model, where the dominant degrees of freedom are the edge states localized along the zigzag boundaries. The resulting model takes the form of a modified Su-Schrieffer-Heeger (SSH)-type ladder with intraedge hopping $t_1$ and $t_2$, and an interedge coupling $t'$ with the effective Hamiltonian as $H(k) = (t_1+t_2)\cos(k)\sigma_0+(t_1-t_2)\cos(k) \sigma_z+t'\sigma_x$, where the Pauli matrices act on the two edge-state components~\cite{ssh-1,ssh-2}.

In contrast to the zigzag edge, the LDOS of the armchair-edge GNH indicates that the electronic states are predominantly distributed over the interior region of the lattice, reflecting the bulk state feature of the corresponding bands. 
The bands shown in Fig.~\ref{fig:armchair_ldos}(a) are therefore mainly governed by the confined bulk states rather than the edge states. Nevertheless, when zigzag-type junction corners appear at the intersection of the armchair edges, localized states can emerge in these regions and contribute to the LDOS of the bands close to the Fermi level. Those types of localization might be the signatures of the higher-order topological insulators at the corner junction; however, that lies beyond the scope of our work since the corresponding calculations are not applicable for the effective 1D system. As illustrated in the LDOS maps, these corner states are spatially confined and only affect the band close to the Fermi energy, while bands far away from the Fermi level remain dominated by the bulk contributions distributed throughout the lattice. This observation suggests that the GNH realizes a type of graphene nanoribbon junctions~\cite{cao-z2-prl17, han-junction-prb23, johan-junction-prb24}. The armchair component is mainly confined to the bulk state, while the zigzag component can support localized edge modes. 

\vspace{0.1in}

\section{Alternating Zak phase} 
\label{sec:zak}

Furthermore, we study the the topological properties of the graphene nanohelicoid, we use the full tight-binding description for the 1D chain on $l^2(\mathbb{Z})$. The momentum-shifted particle-hole relation $E_{v}(k) = -E_{c}(k+\pi)$ is the consequence of the nonsymmorphic symmetry, which can be represented in the momentum space as $\mathcal{S} \mathcal{H}_{k} \mathcal{S}^{-1} = - \mathcal{H}_{k+\pi}$, where $\mathcal{S} $ is the sublattice symmetry operator~\cite{YXZhaonatcomm,PhysRevB.93.195413}. It is known that the gapped systems with this symmetry possess class-II sublattice symmetry and are topologically trivial in one dimension~\cite{YXZhaonatcomm}. However, this bulk topological invariant does not represent the termination-dependent number of boundary states for the finite systems.

Hence, to perform the calculations, we construct chiral lattice model from two triangular sectors ($\times 2$ supercell), presented in Fig.~\ref{fig:spiral_lattice}. In such a system, one can define the strong topological invariant in the ten-fold way classification~\cite{Ryu_2010,RevModPhys.88.035005}; however, one should always keep in mind that the translation operator has been modified and this topological invariant should not be used to characterize the system, described in Sec.~\ref{sec:band}. 
We would start with the zigzag edge, where similarly to Eq.~(\ref{eq:hamiltonian-def-anti}), the atoms are labeled in the way $\left\{ A_{m,l},B_{m,l}\right\}$, describing $m$-pairs of $A,B$ atoms on the $l$-ring of the helicoid. Therefore, the chiral tight-binding Hamiltonian of this system reads as follows:
\begin{widetext}
\begin{equation}
\label{eq:hamiltonian-def-v2}
\begin{aligned}
\hat{H}&= t \sum_{i \in \mathbb{Z}} \Bigg\{ \sum_{l=n}^{N}\sum_{m=1}^{2l-1}  \Big[ \hat{a}^{\dagger}_{i,m,l}\hat{b}_{i,m,l} +  \hat{b}^{\dagger}_{i,m,l}\hat{a}_{i+\floor*{\frac{m}{2l-1}},m-m \floor*{\frac{m}{2l-1}} +1,l}   \Big] \\
&+ \sum_{l=n}^{N-1}\sum_{\left< \right. m_1,m_2 \left. \right>}    \hat{a}^{\dagger}_{i,m_1,l}\hat{b}_{i,m_2,l+1}   + \sum_{l=n}^{N-1} \sum_{\left< \right.  m_1^{\prime},m_2^{\prime} \left. \right>}    \hat{b}^{\dagger}_{i,m_1^{\prime},l}\hat{a}_{i,m_2^{\prime},l+1}     +\text{h.c.} \Bigg\}.
\end{aligned}
\end{equation}
\end{widetext}
Here, the termination for the finite chains defines the bulk unit cell convention.
In general, one may straightforwardly generalize this model for other types of terminations and edges, for example, an armchair-edge helicoid with zigzag-type junctions corners, as presented in Fig.~\ref{fig:armchair_type} (type-II). In such cases, the resulting Hamiltonian satisfies the properties discussed below, and therefore the following formalism is applicable for calculating the topological properties. 

In the Fourier space, the Hamiltonian transforms as follows:
\begin{equation}
\label{eq:hamiltonian-fourier}
\begin{aligned}
\hat{H}= \frac{1}{2\pi}\int_{-\pi}^{\pi} dk\, \Psi_{k}^{\dagger} \mathcal{H}_{k} \Psi_{k},
\end{aligned}
\end{equation}
with $\Psi_{k} = \left( \Psi_{k}^{(a)},\Psi_{k}^{(b)}\right)^{T}$,  
where
\begin{equation}
\label{eq:hamiltonian-fourier-psi}
\begin{aligned}
\Psi_{k}^{(a)}&=\left( \{ \hat{a}_{k,m,n}\}_{m=1}^{2n-1 }, ... \,,\{ \hat{a}_{k,m,N}\}_{m=1}^{2N-1} \right),\\
\Psi_{k}^{(b)}&=\left( \{ \hat{b}_{k,m,n}\}_{m=1}^{2n-1 }, ... \,,\{ \hat{b}_{k,m,N}\}_{m=1}^{2N-1} \right),\\
\end{aligned}
\end{equation}
and the Hamiltonian $ \mathcal{H}_{k} $ is constructed from the following parts:
\begin{equation}
\begin{aligned}
\mathcal{H}_{k} = \left( \begin{array}{cc}
0 & \mathcal{Q}^{\dagger}(k)\\
\mathcal{Q}(k)& 0
\end{array}
\right).
\end{aligned}
\end{equation}
Here, $\mathcal{Q}(k)$ is a block-tridiagonal matrix: 
\begin{equation}
\label{eq:Q-def}
\mathcal{Q}(k) =  \left(\begin{array}{ccc c cc}
h^{(n)} & \Delta_1^{(n)}&  & & \\
\Delta_{2}^{(n)} & h^{(n+1)} &  \Delta_{1}^{(n+1)} & & \\
 &  \Delta_{2}^{(n+1)}  &  \ddots  &  \ddots  & \\
 &   &  \ddots &   \ddots &  \Delta_{1}^{(N-1)} \\
 &   &     & \Delta_{2}^{(N-1)}  & h^{(N)} \\ 
\end{array} \right).
\end{equation}
For $n \le s \le N$, the matrices $h^{(s)}$, $\Delta_1^{(s)}$, and $\Delta_{2}^{(s)}$ are given by
\begin{equation}
\begin{aligned}
\left[h^{(s)}\right]_{ml}&=  t \left[  \delta_{m,l} + \delta_{m+1,l}+ e^{ik} \delta_{m,2s-1}\delta_{l,1} \right],
\end{aligned}
\end{equation}
with $1\le  m \le 2s-1$ and $1\le l \le 2s-1$,
\begin{equation}
\label{eq:delta-1-def}
\begin{aligned}
\left[\Delta_1^{(s)}\right]_{ml}&=  t  \sum_{p=s+1}^{2s}  \delta_{p-1,m}\delta_{ p +1 ,l},
\end{aligned}
\end{equation}
with $1\le  m \le 2s-1$ and $1\le l \le 2s+1$,
\begin{equation}
\label{eq:delta-2-def}
\begin{aligned}
\left[\Delta_2^{(s)}\right]_{ml}&=  t \sum_{p=1}^{s}  \delta_{p,m}\delta_{p ,l},
\end{aligned}
\end{equation}
with $1\le  m \le 2s+1$ and $1\le l \le 2s-1$.

Since $\mathcal{H}_{k}$ possesses the particle-hole, the time-reversal, and the chiral symmetries, it belongs to the BDI class in the ten-fold way classification~\cite{Ryu_2010,RevModPhys.88.035005}. We therefore use these symmetries of $\mathcal{H}_{k}$ to calculate the winding number of the system $\nu$ as follows: 
\begin{equation}
\label{eq:winding-number-def-0}
\begin{aligned}
\nu=\frac{1}{2\pi i} \int_{-\pi}^{\pi} dk \, \text{tr}\left[ \mathcal{Q}^{-1}(k) \partial_{k} \mathcal{Q}(k)\right].
\end{aligned}
\end{equation}

Since for the matrix $U$, the following relation holds $\text{tr} \left( U^{-1} d U \right) = d \left( \log \text{det}\,U\right)$, then after making the substitution $z=e^{ik}$ in Eq.~(\ref{eq:winding-number-def-0}), the winding number is given by: 
\begin{equation}
\label{eq:winding-number-def}
\begin{aligned}
\nu
&=\oint_{|z| = 1} \frac{dz}{2\pi i}  \frac{d  }{ d z} \log  \text{det}\,\mathcal{Q}(z)  \\
&=\sum_{z_a\in A}  \eta(z_a) \underset{z=z_{a}}{\text{res}} \left[  \frac{d  }{ d z} \log  \text{det}\,\mathcal{Q}(z)  \right],
\end{aligned}
\end{equation}
where $A$ is a set of zeros $\text{det}\,\mathcal{Q}(z)  =0$ and $\eta(z_a) =1$ for $|z_a| <1$, $\eta(z_a) =0$ for $|z_a|>1$.

Let us analyze the properties of zeros of $\text{det}\,\mathcal{Q}(z)$. In this case, $\text{det}\,\mathcal{Q}(z)$ is a self-reciprocal polynomial of the order $W=N-n+1$ (see Appendix~\ref{sec:schur-proof}):
\begin{equation}
\label{eq:detQ-equality}
\text{det}\,\mathcal{Q}(z) = z^{W}\text{det}\,\mathcal{Q}(1/z) .
\end{equation}
Hence, $\text{det}\,\mathcal{Q}(z)$ can be written in the following  form: 
\begin{equation}
\label{eq:zero-det}
\text{det}\,\mathcal{Q}(z) = \sum_{i=0}^{W} p_{i} z^{i},\quad p_{i}=p_{W-i},\quad p_{i}  > 0 .
\end{equation}
Therefore, the roots of $\text{det}\,\mathcal{Q}(z) =0$ appear in reciprocal pairs $\left(z^{(k)} , 1/z^{(k)} \right)$ with identical multiplicity $\alpha_{k}$. 
Consequently, Eq.~(\ref{eq:winding-number-def}) can be explicitly simplified as follows:
\begin{equation}
\label{eq:winding-number-simp}
\nu
=\sum_{k} \oint_{|z| = 1} \frac{dz}{2\pi i}  \left[\frac{\alpha_{k}}{z-z^{(k)}}+\frac{\alpha_{k}}{z-1/z^{(k)}}\right]  .
\end{equation}

\begin{figure}[t]
\includegraphics[width=8.6cm]{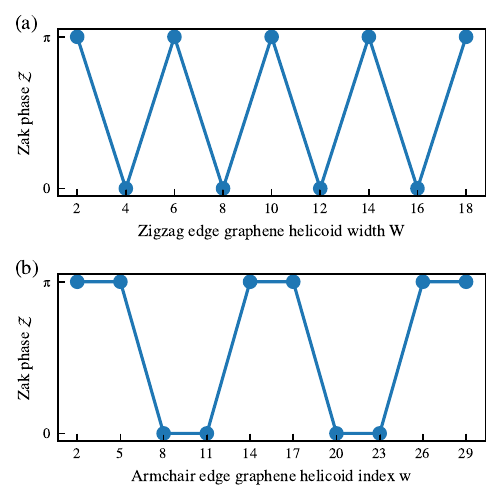}
\caption{The Zak phase evolution. (a) Dependence of the Zak phase on the helicoid width for the zigzag edge GNH. (b) Dependence of the Zak phase on the armchair structure index for the armchair edge GNH.}
\label{fig:zak-phase}
\end{figure}

From Eq.~(\ref{eq:zero-det}), we obtain that $z=-1$ is the solution of $\text{det}\,\mathcal{Q}(z) =0$ for every odd width $W$ (which corresponds to the metallic regime), and hence, the winding number in Eq.~(\ref{eq:winding-number-def}) is not well-defined for this case. Therefore, we focus on only on even $W$ for which $z=-1$ is not a solution (see Appendix~\ref{sec:real-proof} for details). According to Eqs.~(\ref{eq:p_k-recurernce}) and (\ref{eq:q_k-recurernce}), the root multiplicities satisfy $2 \sum_{k}\alpha_{k} = W$. Hence, we obtain that the winding number is given by
\begin{equation}
\label{eq:winding-number-res}
\begin{aligned}
\nu
&=\sum_{k} \alpha_{k} = \frac{W}{2}.
\end{aligned}
\end{equation}
The Zak phase is related to the winding number through $\mathcal{Z} = \pi\nu\pmod{2\pi}$ and retains only the parity of the winding number, which is a $\mathbb Z_2$ reduction of the full $\mathbb Z$ classification. It is given by:
\begin{equation}
\label{eq:zak-phase-expl}
\mathcal{Z}= \pi\left[  \frac{W}{2}\, \left(\text{mod}\, 2\right)\right] = \begin{cases}
\pi, \,\, W =4 m-2, \\
0, \,\, W =4 m ,\\
\end{cases}
\end{equation}
with $m\in \mathbb{N}$ and is presented in Fig.~\ref{fig:zak-phase}(a). Notice that for given convention $z=e^{ik}$ and momentum orientation, the winding number $\nu$ is positive. Reversing the momentum orientation $k \rightarrow -k$ or interchanging the sublattices leads to the sign reversal of $\nu$ while preserving the number of boundary states and the Zak phase.

\begin{figure}[t]
\includegraphics[width=.45\textwidth]{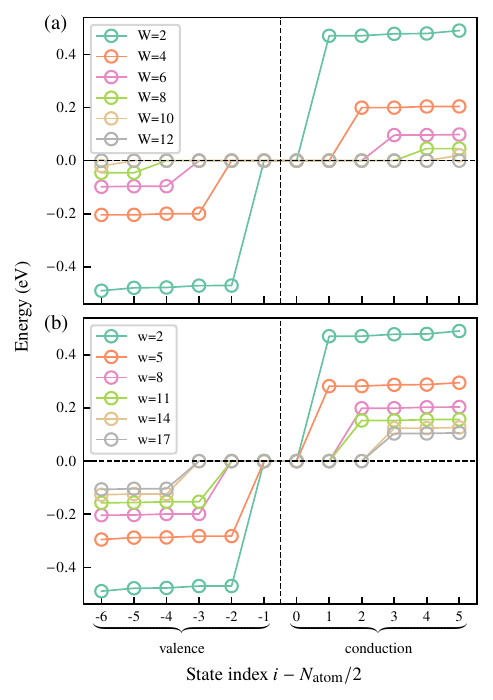}
\caption{Discrete energy spectrum for finite size $L=30$ GNH. The number of atoms in the finite chain is defined as $N_{\text{atom}}$. (a) The zigzag-edge GNH with different widths. (b) The type-II armchair-edge GNH with different armchair structure indices.
}
\label{fig:finite_edge}
\end{figure}
\begin{figure}[t]
	\centering
	\includegraphics[width=9cm]{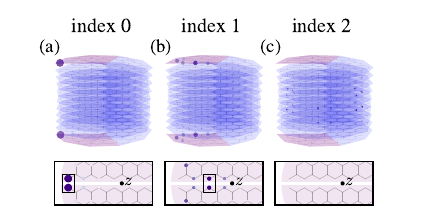}
	\caption{Local density of states corresponding to Fig.~\ref{fig:finite_edge}(a) (width $W=4$) with conduction state index. Panels (a) and (b) represent the two boundary states, and panel (c) represents a bulk state. 
    }    
    \label{fig:edge-states}
\end{figure}

According to the bulk-boundary correspondence, the helicoid width determines the amount of boundary states for the finite chain, which is equal to $W/2$ per boundary. Hence, by considering the finite chain of the length $L$ without making a loop, we obtain the following ordering of the eigenvalues for the boundary states with the conduction state $0 < E_{0} < ... <  E_{\frac{W}{2}-1}$, which is presented in Fig.~\ref{fig:finite_edge}(a),
with
\begin{equation}
\lim_{L\rightarrow \infty} E_{i}=0,\,\,\,\,0\le i \le \frac{W}{2}-1.
\end{equation}
Since the chiral model preserves particle-hole symmetry, the corresponding conduction and valence states exhibit symmetric LDOS.
Each of the boundary states are predominantly localized at the sites $A_{1,1,N-2i}$ and $B_{L,2N-4i-1,N-2i}$ for $E=\pm E_i$.
Thus in the zigzag edge GNH, we obtain one boundary mode per two coupled SSH chains for one open boundary, which can be demonstrated numerically by considering the LDOS for the finite chain. The LDOS of the $W/2$ boundary states is presented in Fig.~\ref{fig:edge-states} for $W=4$. 

\begin{figure}[t]
	\centering
	\includegraphics[width=6.8cm]{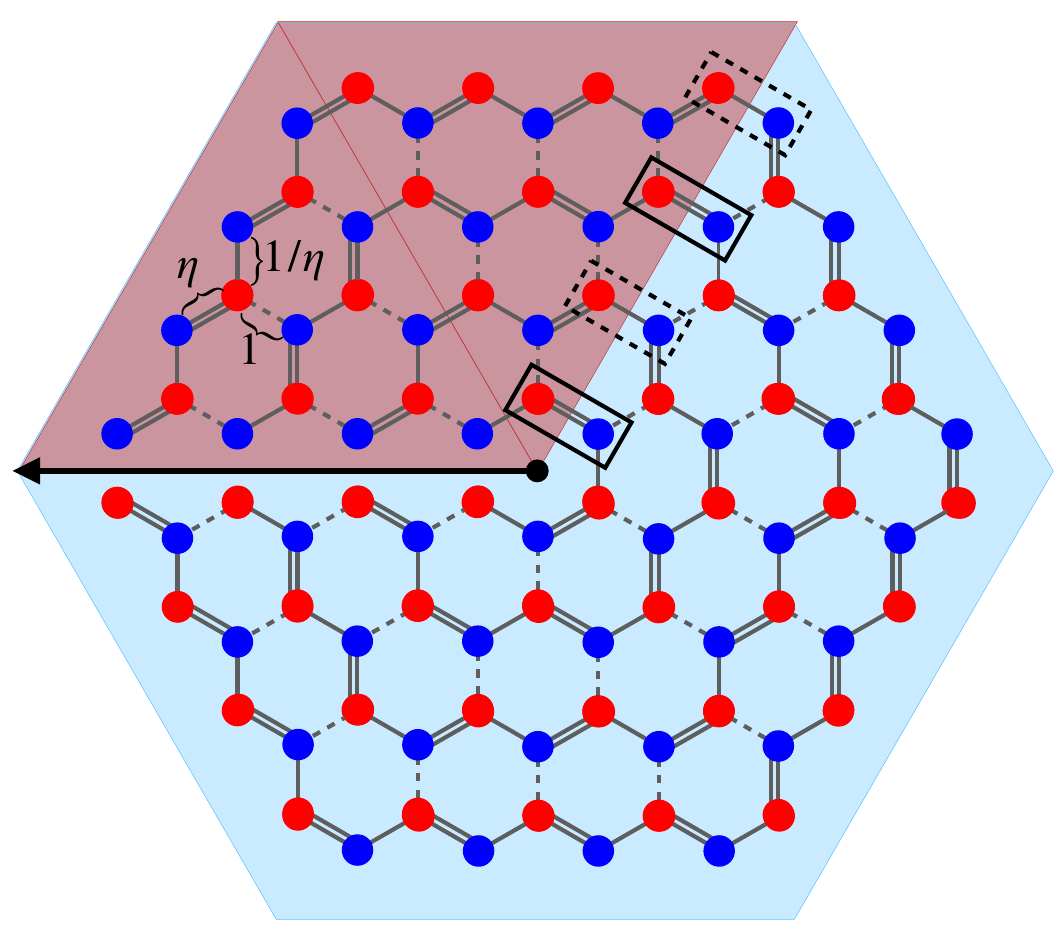}
	\caption{Schematic lattice model of the zigzag-edge GNH with lifted degeneracy of the hopping parameters; the unit cell is highlighted in red (chiral model). Charge localization in the atomic limit when cutting the chain after a full unit cell is presented by solid rectangle ($\eta \rightarrow \infty$) and dashed rectangle ($\eta\rightarrow 0$).
    }
    \label{fig:Spiral_NSSH_lattice}
\end{figure}

In this model, the Zak phase exhibits an alternating behavior depending on the helicoid width $W$. The Zak phase defines the bulk charge polarization and thus determines the positions of the charge center. 
Consequently, the charge center is displaced by exactly half a lattice spacing of the effective 1D chain for $W=4m-2$, and the charge center is localized at the unit cell for $W=4m$.
As was mentioned earlier, the GNH with the zigzag edge can be considered as a set of $W$ SSH chains as presented in Fig.~\ref{fig:spiral_lattice}(b), with the interchain coupling described by the $\Delta_{1,2}$ blocks in Eq.~(\ref{eq:Q-def}). Since $\text{det}\,\mathcal{Q}(z,\eta)$ of the model (\ref{eq:hamiltonian-def-v2}) with unequal hoppings within the chains (see Fig.~\ref{fig:Spiral_NSSH_lattice}) has $W$ real zeros and $z=-1$ is not the root of $\text{det}\,\mathcal{Q}(z,\eta)=0$ for any $\eta \in \mathbb{R}_{+}$ if $W$ is even, then there exists an adiabatic connection between the systems with $\eta=1$ and $\eta \neq 1$ (see Appendix~\ref{sec:real-proof}). 

As a result, the gap does not close as we continuously modify $\eta\in \mathbb{R}_{+}$. Therefore, we can demonstrate the localization of the one boundary mode per two coupled SSH chains by considering the atomic limit either for $\eta \rightarrow \infty$ or $\eta \rightarrow 0$ for the GNH, presented in  Fig.~\ref{fig:Spiral_NSSH_lattice} by solid and dashed rectangles respectively. By counting the resulting half-charges, one can diagrammatically see that indeed the bulk polarization is $P=0$ for $W=4m$ and $P=e/2$ for $W=4m-2$. Hence, the Zak phase is indeed modified only when the system width $W$ is changed with the increment of $2$. 

\begin{figure}[t]
	\centering
	\includegraphics[width=0.48\textwidth]{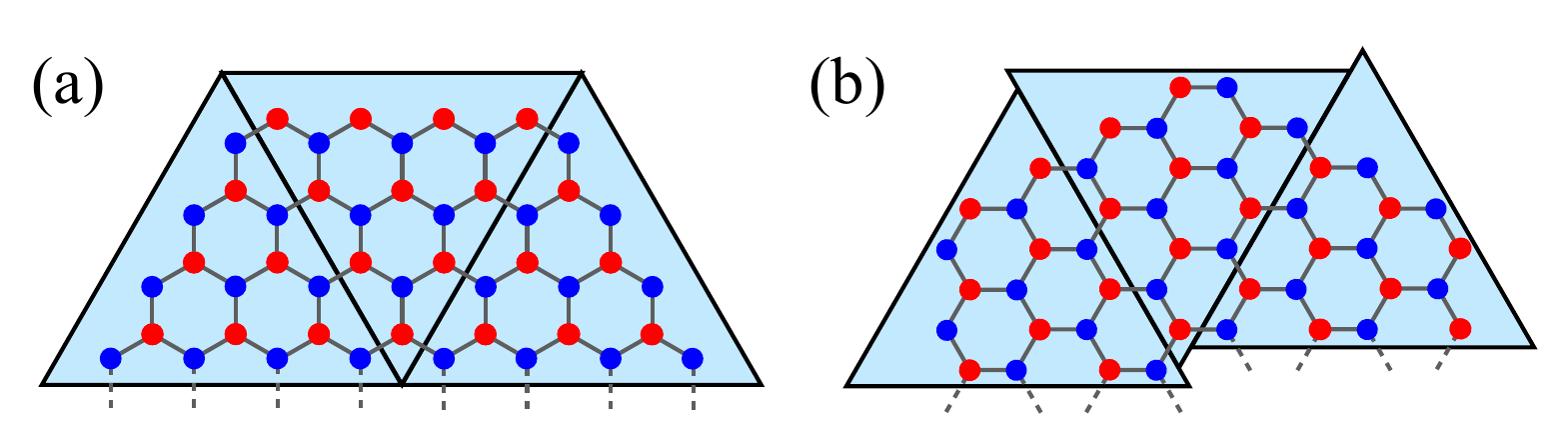}
	\caption{Schematic graph of the infinite chain for a zigzag-edge GNH with two examples of the unit cells determined by the termination (represented by triangles). (a) Unit cell is defined as in Fig.~\ref{fig:spiral_lattice}(c). (b) Unit cell is  defined as a relative $\pi/6$-rotation compared to the unit cell defined in Fig.~\ref{fig:spiral_lattice}(c).}    
    \label{fig:term-types}
\end{figure}

Note that, the existence and such localization of the boundary modes is attributed to this particular definition of the unit cell. This property is a direct consequence of the Toeplitz index theorem applied for such systems~\cite{MonvelToeplitz,10.1063/5.0150870}. 
Indeed, let us denote the number of $A$ and $B$ atoms in the unit cell for the anti-bipartite lattice as $\mathcal{N}_{A}$ and $\mathcal{N}_{B}$. The winding number $\nu$ of the system with doubled unit cell is given by $\nu=\left(\mathcal{N}_{A}-\mathcal{N}_{B}\right)/2$, as will be shown later. With the redefinition of the unit cell, the number of atoms $\mathcal{N}_{A}^{\prime}$ and $\mathcal{N}_{B}^{\prime}$ may change; thus, the winding number $\nu^{\prime}$ is modified. For example, if one redefines the unit cell as a relative $\pi/6$-rotation compared to the unit cell defined in Fig.~\ref{fig:spiral_lattice}(c), the number of atoms $\mathcal{N}_{A}^{\prime}$ and $\mathcal{N}_{B}^{\prime}$ always equal for all system sizes, which results in $\nu^{\prime}=0$ and the absence of the boundary modes for the finite system for this termination type. 
Such redefinition of the unit cells is schematically demonstrated in Fig.~\ref{fig:term-types} for a zigzag-edge GNH with $W=4$.

Indeed, for a 1D nonsymmorphic system, one can make a general statement about the termination-resolved bulk-boundary correspondence and derive a lemma for counting the boundary states with opposite chirality, which is, in particular, applicable to the graphene nanohelicoids. Let us consider a system that satisfies the following  symmetry~\cite{YXZhaonatcomm,PhysRevB.93.195413}:
\begin{equation}
\label{eq:anti-chiral-symmetry}
\mathcal{S} \mathcal{H}_{k} \mathcal{S}^{-1} = - \mathcal{H}_{k+\pi} .
\end{equation}
Therefore, in diagonal basis for the sublattice symmetry operator $\mathcal{S}$:
\begin{equation}
\mathcal{S} = \left( \begin{array}{cc}
\mathbb{I}_{\mathcal{N}_{A}} & 0 \\
0 &  -\mathbb{I}_{\mathcal{N}_{B}}
\end{array} \right),
\end{equation}
the general form of the Hamiltonian is given by
\begin{widetext}
\begin{equation}
\label{eq:H-nonsym}
\mathcal{H}(z) = \left( \begin{array}{cc}
 \sum_{l \in \mathbb{N}} A_{2l-1} z^{2l-1} +\sum_{l \in \mathbb{N}} A_{2l-1}^{\dagger} z^{-2l+1} & \sum_{l \in \mathbb{Z} } Q_{-l}^{\dagger} z^{2l}  \\
\sum_{l \in \mathbb{Z} } Q_{l} z^{2l} &   \sum_{l \in \mathbb{N}} B_{2l-1} z^{2l-1} +\sum_{l \in \mathbb{N}} B_{2l-1}^{\dagger} z^{-2l+1}
\end{array} \right),
\end{equation}
\end{widetext}
where we have made a substitution $z=e^{ik}$ and $A_{2l-1}$ are $\mathcal{N}_{A} \times \mathcal{N}_{A}$ matrices, $B_{2l-1}$ are $\mathcal{N}_{B} \times \mathcal{N}_{B}$ matrices, $Q_{l}$ are $\mathcal{N}_{B} \times \mathcal{N}_{A}$ matrices. The sizes $\mathcal{N}_{A},\mathcal{N}_{B}$ are determined by the unit cell definition.

\textit{Lemma.} The topological properties of the gapped 1D systems with the finite range interaction, such that $\mathcal{S} \mathcal{H}_{k} \mathcal{S}^{-1} = - \mathcal{H}_{k+\pi} $, are determined by the order of self-inverse polynomial\footnote{By definition, a self-inverse polynomial of the order $s$ satisfies \begin{equation*}
p_{s}(\zeta) = \zeta^{s} \overline{p_{s} \left(\frac{1}{\bar{\zeta}} \right)} \end{equation*} } $p(\zeta)$ and the valuation of $\text{det}\,\tilde{Q}(\zeta)$ at $\zeta=0$, i.e., 
\begin{equation}
\text{det}\,\tilde{Q}(\zeta) = \sum_{i=\beta}^{\alpha} c_{i} \zeta^{i} = \zeta^{\beta} p_{\alpha-\beta}(\zeta),
\end{equation}
where $\zeta = z^2$,
the matrix $\tilde{Q}(\zeta)$ is defined as:
\begin{widetext}
\begin{equation}
\label{eq:tilde-Q-nonsym}
\tilde{Q}(\zeta) = \left( \begin{array}{cc}
 \sum_{l \in \mathbb{N}} A_{2l-1} \zeta^{l} +\sum_{l \in \mathbb{N}} A_{2l-1}^{\dagger} \zeta^{1-l} & \sum_{l \in \mathbb{Z} } Q_{-l}^{\dagger} \zeta^{l}  \\
\sum_{l \in \mathbb{Z} } Q_{l} \zeta^{l} &   \sum_{l \in \mathbb{N}} B_{2l-1} \zeta^{l-1} +\sum_{l \in \mathbb{N}} B_{2l-1}^{\dagger} \zeta^{-l}
\end{array} \right),
\end{equation}
\end{widetext}
and $\alpha+\beta=\mathcal{N}_{A}-\mathcal{N}_{B}$. Consequently, the winding number is given by:
\begin{equation}
\label{eq:1d-index-anti-chiral}
\nu = \frac{1}{2}\,\text{tr}\,\mathcal{S}.
\end{equation}

\textit{Proof.} Let us consider the doubled Hamiltonian as $\tilde{\mathcal{H}}_{k} = \text{diag} \left( \mathcal{H}_{k},\mathcal{H}_{k+\pi}\right)$, with corresponding chiral operator $\tilde{\mathcal{S}}$ defined as
\begin{equation}
\tilde{\mathcal{S}} = \left( \begin{array}{cc}
0&\mathcal{S}\\
\mathcal{S}&0
\end{array} \right).
\end{equation}
By diagonalizing $\tilde{\mathcal{S}}$, the doubled Hamiltonian $\tilde{\mathcal{H}}(\zeta)$ is given by:
\begin{equation}
\tilde{\mathcal{H}}(\zeta)= \left( \begin{array}{cc}
0&\tilde{\mathcal{Q}}^{\dagger}(\zeta)\\
\tilde{\mathcal{Q}}(\zeta)&0
\end{array} \right),
\end{equation}
with $\zeta=z^2$ and $\tilde{\mathcal{Q}}$ being given by Eq.~(\ref{eq:tilde-Q-nonsym}). Since the system is gapped:
\begin{equation}
\text{det}\, \mathcal{H}_{k} \neq 0,\,\,\,\,\forall k\in \left[-\pi,\pi\right],
\end{equation}
and by taking the determinant of Eq.~(\ref{eq:anti-chiral-symmetry}),
\begin{equation}
\text{det}\, \mathcal{H}_{k+\pi} = (-1)^{\mathcal{N}_{A}+\mathcal{N}_{B}} \text{det}\, \mathcal{H}_{k},
\end{equation}
$\mathcal{N}_{A}+\mathcal{N}_{B}$ is always an even number.

For the finite range interaction, the sums in $\text{det} \, \tilde{Q}(\zeta)$ are finite. Thus, $\text{det} \, \tilde{Q}(\zeta)$ is a Laurent polynomial, which can be presented as follows:
\begin{equation}
\label{eq:det-tilde-Q-anzatz}
\text{det}\,\tilde{Q}(\zeta) = \zeta^{\beta} p_{\alpha-\beta}(\zeta).
\end{equation}
Let us consider the auxiliary function $D(\zeta)$ defined as
\begin{equation}
D(\zeta) = \text{det}\left[U(\zeta) \tilde{Q}(\zeta) V(\zeta) \right],
\end{equation}
with
\begin{equation}
U(\zeta)=\left( \begin{array}{cc}
 \mathbb{I}_{\mathcal{N}_{A}}\zeta^{-1/2} & 0 \\
0 &  \mathbb{I}_{\mathcal{N}_{B}}
\end{array} \right),\,\,\,\,V(\zeta)  = \left( \begin{array}{cc}
\mathbb{I}_{\mathcal{N}_{A}} & 0 \\
0 &  \mathbb{I}_{\mathcal{N}_{B}} \zeta^{1/2}
\end{array} \right).
\end{equation}
Thus, $D(\zeta)$ is given by
\begin{equation}
D(\zeta) = \zeta^{\frac{\mathcal{N}_{B}-\mathcal{N}_{A}}{2}}\text{det}\left[ \tilde{Q}(\zeta)  \right]
\end{equation}
and it satisfies the following property:
\begin{equation}
\label{eq:D-property}
\overline{D(1/\zeta)} =  D (\bar{\zeta}).
\end{equation}
By substituting Eq.~(\ref{eq:det-tilde-Q-anzatz}) into Eq.~(\ref{eq:D-property}), we obtain that
\begin{equation}
p_{\alpha -\beta}(\zeta) = \zeta^{\mathcal{N}_{A}-\mathcal{N}_{B}-2\beta} \overline{ p_{\alpha -\beta} (1/\bar{\zeta})}
\end{equation}
or $p(\zeta)$ is a self-inverse polynomial of the order $\alpha - \beta = \mathcal{N}_{A}-\mathcal{N}_{B}-2\beta$. Therefore, the winding number reads
\begin{equation}
\nu = \frac{\alpha-\beta}{2} +\beta = \frac{\mathcal{N}_{A}-\mathcal{N}_{B}}{2}
\end{equation}
or $\nu =  \text{tr}\,\mathcal{S}/2$ which proves the lemma.

One must specify that it is not a strong bulk invariant, but rather termination-resolved Fredholm index, since the sublattice symmetry of the system is class-II. Hence, the unit cell should be considered as a parametrization of the periodic bulk embedding on a particular geometry of a system that may explicitly depend on the boundary. In general, $\nu$ is the index that fixes the signed imbalance $n_{+}-n_{-}$ of zero modes with opposite chiralities. For the graphene helicoid case, $Q(z=0)$ is invertible. That eliminates opposite-chirality kernel thus making $\nu$ an exact zero mode count for a semi-infinite chain

Similarly, one can consider the gapped type-II armchair-edge case and calculate the corresponding winding number of the Hamiltonian that is given by
\begin{equation}
\label{eq:winding-armchair}
\nu = \floor*{\frac{w^{II}+4}{6}}.
\end{equation}
Indeed, combinatorially for the type-II armchair edge GNH with the unit cell defined as in Fig.~\ref{fig:spiral_lattice}(d), the difference between the numbers of $A$ and $B$ sites is given by:
\begin{equation}
\mathcal{N}_{A}\left(w^{II}\right) - \mathcal{N}_{B}\left(w^{II}\right)= 2 \left( \frac{w^{II}+1}{3}-\floor*{\frac{w^{II}+1}{6}} \right).
\end{equation}
Using Eq.~(\ref{eq:1d-index-anti-chiral}) and $w^{II}  \equiv 2 \pmod{3}$, we obtain the winding number (\ref{eq:winding-armchair}).

The Zak phase dependence on the armchair-edge GNH index $w$ is presented in Fig.~\ref{fig:zak-phase}(b). Consequently, by considering the finite chain of the length $L$, we obtain the following ordering of the eigenvalues for the boundary states for both types presented in Fig.~\ref{fig:finite_edge}(b). This signifies that the armchair-edge localization may involve LDOS distribution predominantly on two atoms instead of on one atom as in zigzag edge. Thus, the boundary modes form a much more complex sequence of localized states. Similarly to the zigzag-edge GNH, the redefinition of the unit cell and corresponding termination changes the number of boundary states.

\section{Discussion and conclusions}
\label{sec:discussion}

In this work, we have shown that the electronic structure of finite width graphene helicoid can be understood from an effective one-dimensional lattice description that directly reflects the underlying geometry. Within this framework, the system exhibits an antichiral (or antibipartite) symmetry, which enforces a momentum-shifted spectral relation $E_v(k)=-E_c(k+\pi)$ if the time-reversal symmetry is present.
A key result is that the finite width of the helicoid acts as a geometric control parameter for the electronic spectrum. The related features follow directly from the lattice structure and its quantization condition, and can be understood analytically within the tight-binding description. 

As the number of rings varies, the zigzag-edge system undergoes repeated gap closing and reopening, leading to the continuous alternation between semiconducting and metallic behaviors. As the width of the gapped system increases, the dispersion becomes flatter, while the low-energy modes become strongly localized along the zigzag edge. In contrast, the armchair-edge system demonstrates different periodicity of the gap closing and reopening depending on the type of the system, as presented in Fig.~\ref{fig:armchair_type}, being gapped only for type-II configurations. In this case, the system exhibits confined bulk-state features, while junctions between two armchair-edge flakes can introduce zigzag-edge segments that give rise to the edge modes observed in the LDOS. 

At the same time, the change of the system size for gapped zigzag-edge and armchair-edge GNHs is accompanied by the alternation of the Zak phase between trivial and nontrivial values (see Fig.~\ref{fig:zak-phase}) for the semiconducting case depending on the unit cell definition. The unit cell convention can be understood as parametrization of the embedding of graphene sheet on the helicoidal surface depending on the termination. The winding number of the chiral Hamiltonian determines that the finite system has $\nu$ boundary states that are localized in a specific way. For the zigzag-edge GNH with the unit cell convention presented in Fig.~\ref{fig:spiral_lattice}(c), $W/2$ boundary states are localized on every second atom of the boundary starting from the outer radius, while for the type-II armchair-edge GNH with the unit cell convention presented in Fig.~\ref{fig:spiral_lattice}(d) localization atoms form a sequence of $\floor*{(w^{II}+4)/6}$ boundary states, as described in Sec.~\ref{sec:zak}. In general, the termination-resolved index of a 1D system with the nonsymmorphic symmetry $\mathcal{S} \mathcal{H}_{k} \mathcal{S}^{-1} = - \mathcal{H}_{k+\pi}$ is determined by the trace of the sublattice symmetry operator $\text{tr}\, \mathcal{S}/2$ or explicitly by the sublattice imbalance for the given termination.

More broadly, this work highlights extended lattice defects as a viable platform for engineering tunable topological electronic states, opening new possibilities for geometry-driven design in graphene systems as a natural extension of the graphene nanoribbon geometry.

\begin{acknowledgments}
We thank Hanning Zhang, Johan Félisaz, Zolt\'{a}n Guba, and Titus Neupert for the fruitful discussions. We acknowledge support by the Swiss National Science Foundation (Grants No.~204254 and No.~224624). Computations were performed at the facilities of the Scientific IT and Application Support Center of EPFL.
\end{acknowledgments}

\appendix

\section{Derivation of Eq.~(\ref{eq:detQ-equality})}
\label{sec:schur-proof}

Let us consider the determinant of the matrix $\mathcal{Q}(z) \equiv \mathcal{Q}_{n}^{N}(z)$ that can be presented using the Schur complement as follows: 
\begin{equation}
\label{eq:proof-1}
\begin{aligned}
\text{det}\,\mathcal{Q}_{n}^{N}(z) = \text{det}\,h^{(n)}(z)\, \text{det} \left[ \tilde{\mathcal{Q}}_{n+1}^{N}(z) \right],
\end{aligned}
\end{equation}
where 
\begin{equation}
\begin{aligned}
\tilde{\mathcal{Q}}_{n+1}^{N}(z)&=\mathcal{Q}_{n}^{N}(z) / h^{(n)}(z)  \\
&= \mathcal{Q}_{n+1}^{N}(z) - \Delta_2^{(n)} \left[ h^{(n)}(z) \right]^{-1} \Delta_1^{(n)} .
\end{aligned}
\end{equation}
Since $\mathcal{Q}_{n}^{N}(z)$ is a block-tridiagonal matrix, only $h^{(n+1)}(z)$ block of the matrix $\mathcal{Q}_{n+1}^{N}(z)$ would be modified while calculating the Schur complement $\mathcal{Q}_{n}^{N}(z) / h^{(n)} (z)$:
\begin{equation}
\label{eq:Schur-complement-h-tilde}
\begin{aligned}
\tilde{h}^{(n+1)}(z)&= h^{(n+1)}(z) - \Delta_2^{(n)} \left[ \tilde{h}^{(n)}(z) \right]^{-1} \Delta_1^{(n)} .
\end{aligned}
\end{equation}
By utilizing Eq.~(\ref{eq:Schur-complement-h-tilde}) in the Schur's formula during the sequential contractions: 
\begin{equation}
\label{eq:schur-comp-consec}
\begin{aligned}
\tilde{\mathcal{Q}}_{n+k+1}^{N}(z)&=\mathcal{Q}_{n+k+1}^{N}(z) / \tilde{h}^{(n+k)}(z)  \\
&= \mathcal{Q}_{n+k+1}^{N}(z) - \Delta_2^{(n+k)} \left[ \tilde{h}^{(n+k)}(z) \right]^{-1} \Delta_1^{(n+k)} ,
\end{aligned}
\end{equation}
we obtain that the determinant of $\mathcal{Q}(z)$ can be rewritten as the following product:
\begin{equation}
\label{eq:Q-det-h-tilde}
\begin{aligned}
\text{det}\,\mathcal{Q}(z) = \prod_{k=0}^{N-n} \text{det} \,\tilde{h}^{(n+k)}(z).
\end{aligned}
\end{equation}
In general, it is obvious from the structure of $h^{(j)}(z)$ that the determinant $\tilde{h}^{(n+k)}(z)$ reads as follows:
\begin{equation}
\label{eq:tilde-h-det-gen}
\begin{aligned}
\text{det}\,\tilde{h}^{(n+k)}(z)&= \frac{p_{k+1}(z)}{p_{k}(z)},
\end{aligned}
\end{equation}
where $p_{k}(z)$ is a polynomial of the order $k$ and we set $t=1$ for simplicity of notations.
By calculating the determinant of $\tilde{h}^{(n+k+1)}(z)$ using Eqs.~(\ref{eq:Schur-complement-h-tilde}) and (\ref{eq:schur-comp-consec}), one can obtain the following property:
\begin{equation}
\label{eq:det-tilde-h-explicitly-A}
\begin{aligned}
\text{det}\, \tilde{h}^{(n+k+1)}(z)&= \left(1+z\right)  + z  \sum_{i,j}(-1)^{n+k+i+j-1} \\
&\times \left\{\Delta_2^{(n+k)} \left[ \tilde{h}^{(n+k)}(z) \right]^{-1} \Delta_1^{(n+k)} \right\}_{ij}  \\
&=  \left(1+z\right)  + z  \sum_{i,j} A^{(n+k)}_{ij} (z) ,\\
\end{aligned}
\end{equation}
where $(-1)^{n+k+i+j-1} A^{(n+k)}_{ij} (z)$, $1\le i,j\le \left(n+k\right)$, is a nonzero block matrix in $\Delta_2^{(n+k)} \left[ \tilde{h}^{(n+k)}(z) \right]^{-1} \Delta_1^{(n+k)}$. From Eq.~(\ref{eq:tilde-h-det-gen}), we obtain that the function $A^{(n+k)}_{ij} (z)$ can be represented as the fraction of two polynomials:
\begin{equation}
A^{(n+k)}_{ij} (z) = \frac{q_k^{(ij)}(z)}{p_{k+1}(z)},
\end{equation}
where $q_k^{(ij)}(z)$ is a polynomial of the order k.

Let us show by induction that $A^{(n+k)}_{ij} (z)$ is a persymmetric matrix
\begin{equation}
\begin{aligned}
A^{(n+k)}_{ij} (z) = A^{(n+k)}_{n+k-j+1,n+k-i+1} (z),
\end{aligned}
\end{equation}
which satisfies
\begin{equation}
\label{eq:A-property-reciprocal}
\begin{aligned}
p_{k+1}(z) A^{(n+k)}_{ij} (z) = z^{k} p_{k+1}\left( \frac{1}{z} \right) A^{(n+k)}_{ji} \left(\frac{1}{z} \right).
\end{aligned}
\end{equation}
or equivalently that the polynomial $\sum_{ij} q_{k}^{(ij)}(z)$ is a self-reciprocal polynomial. In the case of $k=0$, the relation is satisfied since:
\begin{equation}
p_{1}(z)=1+z,\,\,\,\,\,\,A^{(n)}_{ij} (z) =\frac{1}{1+z},\,\,1\le i,j\le n ,\\
\end{equation}
Let us assume that the statement (\ref{eq:A-property-reciprocal}) holds for $A^{(n+k)}_{ij} (z)$. By explicitly calculating $A^{(n+k+1)}_{ij} (z)$, we obtain that:
\begin{equation}
\label{eq:A-induction-step}
\begin{aligned}
A^{(n+k+1)}_{ij} (z)&= \frac{p_{k+1}(z)}{p_{k+2}(z)} \Bigg\{ \Bigg[ 1+ \sum_{i_1=i}^{n+k} \sum_{j_1=1}^{j-1} A_{i_1 j_1}^{(n+k)}(z) \Bigg] \\
&\times \Bigg[ 1+ z \sum_{i_2=1}^{i-1} \sum_{j_2=j}^{n+k} A_{i_2 j_2}^{(n+k)}(z) \Bigg] \\
&-  z \sum_{i_3,j_3=1}^{i-1} \sum_{j_3=1}^{j-1} A_{i_3 j_3}^{(n+k)}(z)  \sum_{i_4=i}^{n+k} \sum_{j_4=j}^{n+k} A_{i_4 j_4}^{(n+k)}(z)   \Bigg\}.
\end{aligned}
\end{equation}
From Eqs.~(\ref{eq:det-tilde-h-explicitly-A}) and (\ref{eq:A-induction-step}), we obtain that the recurrence relation for the polynomials $p_{k}(z)$ and $q_{k}^{(ij)}(z)$ are given by
\begin{equation}
\label{eq:p_k-recurernce}
p_{k+1}(z)
= \left(1+z\right) p_{k}(z) + z  \sum_{i,j=1}^{n+k-1} q^{(ij)}_{k-1} (z) ,\\
\end{equation}
\begin{widetext}
\begin{equation}
\label{eq:q_k-recurernce}
\begin{aligned}
\hspace{-0.15cm} q_{k}^{(ij)} (z)=p_{k}(z) \Bigg\{ \Bigg[ 1+ \sum_{i_1=i}^{n+k-1} \sum_{j_1=1}^{j-1} \frac{q^{(i_1 j_1)}_{k-1}(z)} {p_{k}(z)} \Bigg] \Bigg[ 1+ z \sum_{i_2=1}^{i-1} \sum_{j_2=j}^{n+k-1} \frac{q^{(i_2 j_2)}_{k-1}(z)}{p_{k}(z)} \Bigg]  -  z \sum_{i_3=1}^{i-1} \sum_{j_3=1}^{j-1}  \frac{q^{(i_3 j_3)}_{k-1}(z)} {p_{k}(z)} \sum_{i_4=i}^{n+k-1} \sum_{j_4=j}^{n+k-1}  \frac{q^{(i_4 j_4)}_{k-1}(z)} {p_{k}(z)}  \Bigg\},
\end{aligned}
\end{equation}
with $k \in \mathbb{N} \cup \left\{ 0 \right\}$, $p_{0}(z)=q_{0}^{(ij)}(z)=1$, and $q_{-1}(z)=0$. Note that in Eq.~(\ref{eq:q_k-recurernce}):
\begin{equation}
\begin{aligned}
 \sum_{i_1=i}^{n+k-1} \sum_{j_1=1}^{j-1} q^{(i_1 j_1)}_{k-1}(z)\sum_{i_2=1}^{i-1} \sum_{j_2=j}^{n+k-1} q^{(i_2 j_2)}_{k-1}(z)
 - \sum_{i_3=1}^{i-1} \sum_{j_3=1}^{j-1}  q^{(i_3 j_3)}_{k-1}(z) \sum_{i_4=i}^{n+k-1} \sum_{j_4=j}^{n+k-1}  q^{(i_4 j_4)}_{k-1}(z) \propto p_{k}(z) .
\end{aligned}
\end{equation}
The persymmetricity of the matrix $A^{(n+k+1)}_{ij} (z)$ is obvious:
\begin{equation}
\begin{aligned}
A^{(n+k+1)}_{n+k-j+2,n+k-i+2} (z)
&=  \frac{p_{k+1}(z)}{p_{k+2}(z)} \Bigg\{ \Bigg[ 1+ \sum_{i_1=1}^{j-1} \sum_{j_1=i}^{n+k} A_{n+k-i_1+1,n+k-j_1+1}^{(n+k)}(z) \Bigg] \Bigg[ 1+ z \sum_{i_2=j}^{n+k} \sum_{j_2=1}^{i-1} A_{n+k-i_2+1,n+k-j_2+1}^{(n+k)}(z) \Bigg]  \\
&-  z \sum_{i_3=j}^{n+k} \sum_{j_3=i}^{n+k} A_{n+k-i_3+1,n+k-j_3+1}^{(n+k)}(z)  \sum_{i_4=1}^{j-1} \sum_{j_4=1}^{i-1} A_{n+k-i_4+1,n+k-j_4+1}^{(n+k)}(z)   \Bigg\}\\
&= \frac{p_{k+1}(z)}{p_{k+2}(z)} \Bigg\{ \Bigg[ 1+ \sum_{i_1=1}^{j-1} \sum_{j_1=i}^{n+k} A_{j_1i_1}^{(n+k)}(z) \Bigg] \Bigg[ 1+ z \sum_{i_2=j}^{n+k} \sum_{j_2=1}^{i-1} A_{j_2 i_2}^{(n+k)}(z) \Bigg]  \\
&-  z \sum_{i_3=j}^{n+k} \sum_{j_3=i}^{n+k} A_{j_3 i_3}^{(n+k)}(z)  \sum_{i_4=1}^{j-1} \sum_{j_4=1}^{i-1} A_{j_4 i_4}^{(n+k)}(z)   \Bigg\} = A^{(n+k+1)}_{ij} (z).\\
\end{aligned}
\end{equation}
In addition, since by assumption the property (\ref{eq:A-property-reciprocal}) holds for $A^{(n+k)}_{ij} (z)$, the polynomial $\sum_{ij} q_{k}^{(ij)}(z)$ is a self-reciprocal polynomial by this assumption, and hence the polynomial $p_{k+1}(z)$ is also a self-reciprocal by the recurrence relation (\ref{eq:p_k-recurernce}). Therefore,
\begin{equation}
\begin{aligned}
 z^{k+1} p_{k+2}\left(\frac{1}{z} \right)  A^{(n+k+1)}_{ji} \left(\frac{1}{z} \right)
&=z^{k+1} p_{k+1}\left(\frac{1}{z} \right) \Bigg\{ \Bigg[ 1+ \sum_{i_1=j}^{n+k} \sum_{j_1=1}^{i-1} A_{i_1 j_1}^{(n+k)}\left(\frac{1}{z} \right) \Bigg] \Bigg[ 1+ \frac{1}{z} \sum_{i_2=1}^{j-1} \sum_{j_2=i}^{n+k} A_{i_2 j_2}^{(n+k)}\left(\frac{1}{z} \right) \Bigg]  \\
&-  \frac{1}{z} \sum_{i_3=1}^{j-1} \sum_{j_3=1}^{i-1} A_{i_3 j_3}^{(n+k)}\left(\frac{1}{z} \right)  \sum_{i_4=j}^{n+k} \sum_{j_4=i}^{n+k} A_{i_4 j_4}^{(n+k)}\left(\frac{1}{z} \right)   \Bigg\}\\
&=  p_{k+1}(z) \Bigg\{  \Bigg[ 1+ z \sum_{i_1=j}^{n+k} \sum_{j_1=1}^{i-1} A_{j_1 i_1}^{(n+k)}\left(z\right) \Bigg] \Bigg[ 1+ \sum_{i_2=1}^{j-1} \sum_{j_2=i}^{n+k} A_{j_2 i_2}^{(n+k)}\left(z\right) \Bigg]  \\
&- z \sum_{i_3=1}^{j-1} \sum_{j_3=1}^{i-1} A_{j_3 i_3}^{(n+k)}\left(z \right)  \sum_{i_4=j}^{n+k} \sum_{j_4=i}^{n+k} A_{j_4 i_4}^{(n+k)}\left(z \right)   \Bigg\} = p_{k+2}(z) A^{(n+k+1)}_{ij} \left( z \right),\\
\end{aligned}
\end{equation}
which completes the induction step. Hence, $A^{(n+k)}_{ij} (z)$ is a persymmetric matrix that satisfies Eq.~(\ref{eq:A-property-reciprocal}) and as a result, the polynomials $\sum_{ij} q_{k}^{(ij)}(z)$ and $p_{k}(z)$ are the self-reciprocal polynomials.

By combining these results and using Eq.~(\ref{eq:Q-det-h-tilde}), we finally obtain:
\begin{equation}
\text{det}\,\mathcal{Q}(z) =p_{W}(z),
\end{equation}
which proves Eq.~(\ref{eq:detQ-equality}).

If one considers the model with lifted degeneracy described by $\mathcal{Q}(z,\eta)$ as presented in Fig.~\ref{fig:Spiral_NSSH_lattice}, the matrix $A_{ij}^{(n+k)}(z,\eta)$ remains persymmetric; however, the recurrence relations for the resulting polynomials would be modified as follows:
\begin{equation}
\label{eq:p_k-recurernce-eta}
p_{k+1}^{(n)}(z,\eta)
= p_{k}^{(n)}(z,\eta) p_{1}^{(n+k)}(z,\eta^{\alpha_k}) + z  \sum_{i,j=1}^{n+k-1} q^{(ij)}_{k-1,n} (z,\eta) \eta^{2\alpha_k(j-i)} ,\\
\end{equation}
\begin{equation}
\begin{aligned}
\label{eq:q_k-recurernce-eta}
q_{k,n}^{(ij)} (z,\eta)&=\eta^{2 \alpha_k(i-j)} p_{k}^{(n)}(z,\eta) \Bigg\{ 1 + \eta^{2\alpha_k(n+k)- \alpha_k}  \sum_{i_1=i}^{n+k-1} \sum_{j_1=1}^{j-1} \frac{q^{(i_1 j_1)}_{k-1,n}(z,\eta) \eta^{2\alpha_k(j_1-i_1)}} {p_{k}^{(n)}(z,\eta)}  \\
&+  z\, \eta^{-2\alpha_k(n+k)+\alpha_k}  \sum_{i_2=1}^{i-1} \sum_{j_2=j}^{n+k-1} \frac{q^{(i_2 j_2)}_{k-1,n}(z,\eta) \eta^{2\alpha_k(j_2-i_2)}}{p_{k}^{(n)}(z,\eta)}  \\
&+z \sum_{i_1=i}^{n+k-1} \sum_{j_1=1}^{j-1} \frac{q^{(i_1 j_1)}_{k-1,n}(z,\eta) \eta^{2\alpha_k(j_1-i_1)}} {p_{k}^{(n)}(z,\eta)} \sum_{i_2=1}^{i-1} \sum_{j_2=j}^{n+k-1} \frac{q^{(i_2 j_2)}_{k-1,n}(z,\eta) \eta^{2\alpha_k(j_2-i_2)}}{p_{k}^{(n)}(z,\eta)} \\
&-  z \sum_{i_3=1}^{i-1} \sum_{j_3=1}^{j-1}  \frac{q^{(i_3 j_3)}_{k-1,n}(z,\eta)  \eta^{2\alpha_k(j_3-i_3)} }{p_{k}^{(n)}(z,\eta)} \sum_{i_4=i}^{n+k-1} \sum_{j_4=j}^{n+k-1}  \frac{q^{(i_4 j_4)}_{k-1,n}(z,\eta) \eta^{2\alpha_k(j_4-i_4)} } {p_{k}^{(n)}(z,\eta)}  \Bigg\},
\end{aligned}
\end{equation}
\end{widetext}
where due to $\eta$-dependence of the blocks $\tilde{h}^{(n+k)}(z,\eta)$, polynomials acquire dependence on
the size of these blocks. The discrete parameter $\alpha_k$ satisfies $\alpha_k=(-1)^{k}$. Here, the initial polynomials are given by:
\begin{equation}
\begin{aligned}
p_{1}^{(n)}(z,\eta)&= \eta^{2n-1} + \frac{z}{\eta^{2n-1}} ,\\
q_{0,n}^{(ij)}(z,\eta)&=\eta^{2(i-j)},\,\,1\le i,j \le n.
\end{aligned}
\end{equation}
By construction [due to the symmetries of $\mathcal{Q}(z,\eta)$] of the polynomials $p_{k}^{(n)}(z,\eta)$, we obtain that all roots of $\det\,\mathcal{Q}(z,\eta)$ are real and $p_{2k}^{(n)}(-1,\eta)\neq 0$ for every $k\in \mathbb{N}$ and $\eta \in \mathbb{R}_{+}$ (see Appendix~\ref{sec:real-proof} for details).

\section{Nature of roots and adiabatic connectivity}
\label{sec:real-proof}

First, let us consider the roots of $\text{det}\, \mathcal{Q}(z,\eta)$ for arbitrary $\eta >0$ and demonstrate that they are real and negative. Indeed, $\mathcal{Q}(z,\eta)$ is a block-tridiagonal matrix: 
\begin{equation}
\label{eq:Q-eta-def}
\mathcal{Q}(z,\eta) =  \left(\begin{array}{ccc c cc}
h^{(n)}(z,\eta) & \Delta_1^{(n)}&  &  \\
\Delta_{2}^{(n)} & h^{(n+1)}(z,\eta) &  \ddots& & \\
   &  \ddots  &  \ddots  &  \Delta_{1}^{(N-1)} \\
    &     & \Delta_{2}^{(N-1)}  & h^{(N)}(z,\eta) \\ 
\end{array} \right),
\end{equation}
where
\begin{equation}
\begin{aligned}
\left[h^{(s)}(z,\eta)\right]_{ml}&=  \eta^{\alpha_{s}}  \delta_{m,l} + \eta^{-\alpha_{s}} \delta_{m+1,l} \\
&+ \eta^{-\alpha_{s}}  z \delta_{m,2s-1}\delta_{l,1} ,
\end{aligned}
\end{equation}
with $1\le  m \le 2s-1$, $1\le l \le 2s-1$, and $n \le s \le N$. The matrices $\Delta_1^{(s)}$ and $\Delta_{2}^{(s)}$ are defined as in Eqs.~(\ref{eq:delta-1-def}) and (\ref{eq:delta-2-def}), with $t=1$. Hence, the polynomial $p_{W}^{(n)}(z,\eta)$ can be represented as follows:
\begin{equation}
\label{eq:p-real-def}
p_{W}^{(n)}(z,\eta) =  \text{det} \left[ \mathcal{Q}(0,\eta)\right] \text{det} \left[ \mathbb{I} + z \, \hat{\eta}  B(\eta) \right],
\end{equation}
where we have introduced the following notations:
\begin{equation}
B(\eta) =\,\mathbb{E}^{T} \mathcal{Q}^{-1} (0,\eta) \mathbb{J} \mathbb{E},\\
\end{equation}
and
\begin{equation}
\begin{aligned}
\mathbb{J} &=\text{diag} \left( \mathbb{J}^{(n)}, \mathbb{J}^{(n+1)}, ..., \mathbb{J}^{(N)}\right),\\
\mathbb{E} &= \text{diag} \left( e_{1}^{(n)} , e_{1}^{(n+1)}, ..., e_{1}^{(N)}\right), \\
\hat{\eta} & = \text{diag} \left( \eta^{-\alpha_{1}}  , \eta^{-\alpha_{2}},...,\eta^{-\alpha_{W}} \right),
\end{aligned}
\end{equation}
with $\mathbb{J}^{(s)}$ being $(2s-1) \times (2s-1)$ exchange matrix, $e^{(s)}_{1}= (1,0,...,0)^{T}$ being $(2s-1)$-component vector. Since the matrices $h^{(s)}(0,\eta)$ and $\Delta_{1,2}^{(s)}$ satisfy
\begin{equation}
\begin{aligned}
\mathbb{J}^{(s)} \left[ h^{(s)}(0,\eta) \right]^{T} \mathbb{J}^{(s)} &= h^{(s)}(0,\eta),\\
\mathbb{J}^{(s)} \left[ \Delta_{2}^{(s)} \right]^{T} \mathbb{J}^{(s+1)} &= \Delta_{1}^{(s)},\\
\mathbb{J}^{(s+1)} \left[ \Delta_{1}^{(s)} \right]^{T} \mathbb{J}^{(s)} &= \Delta_{2}^{(s)},
\end{aligned}
\end{equation}
then 
\begin{equation}
\mathbb{J} \mathcal{Q}^{T} (0,\eta) \mathbb{J} =  \mathcal{Q} (0,\eta) ,
\end{equation}
or equivalently that $\mathcal{Q}^{-1} (0,\eta) \mathbb{J} $ is a symmetric matrix. Therefore, the matrix $\hat{\eta}B(\eta)$ has $W$ real eigenvalues $\left\{ \lambda_{k} \right\}$ and Eq.~(\ref{eq:p-real-def}) can be rewritten as follows:
\begin{equation}
p_{W}^{(n)}(z,\eta) =  \text{det} \left[ \mathcal{Q}(0,\eta)\right] \prod_{k=1}^{W} \left( 1 + z \lambda_{k} \right).
\end{equation}
Let us explore the Sylvester's criterion for the matrix $B(\eta)$. For an invertible $N \times N$ matrix $A$ and increasing index sets $I,J$ of the same cardinality, the following Jacobi's complementary-minor formula holds:
\begin{equation}
\text{det}\, A[I,J] = (-1)^{\sum_{i\in I}i+\sum_{j\in J}j} \text{det}\, A \,\text{det}\, A^{-1}[J^{c},I^{c}] .
\end{equation}
Therefore, the determinants of the leading principal minors of $B(\eta)$ can be represented via $\mathcal{Q}^{-1} (0,\eta)$ as $M_r = \left[\mathcal{Q}^{-1} (0,\eta) \right][I_r,J_r]$, with $r=1,...,W$, and are given by
\begin{equation}
\text{det}\,M_r = \text{det}\left[\mathcal{Q}^{-1} (0,\eta) \right] \text{det}\left\{ \left[\mathcal{Q} (0,\eta) \right] [J^{c}_r,I^{c}_r] \right\},
\end{equation}
with the index sets $\left( I_r,J_r\right)$ being constructed as follows:
\begin{equation}
\begin{aligned}
I_r& = \left\{ i_{a} \right\}_{a=1}^{r} ,\,\,\,\,i_{a} = 1 +\sum_{b=1}^{a-1}\left(2n+2b-3 \right),\\
J_r& = \left\{ j_{a} \right\}_{a=1}^{r} ,\,\,\,\,j_{a} = i_{a} +2 n+2 k -4 .\\
\end{aligned}
\end{equation}
Since by direct calculation
\begin{equation}
\begin{aligned}
\text{det}\left[\mathcal{Q} (0,\eta) \right] &= \prod_{s=n}^{N} \eta^{(2s-1) \alpha_{s}},\\
\text{det} \left\{ \left[\mathcal{Q} (0,\eta) \right][J^{c}_r,I^{c}_r] \right\} &= \frac{ \prod_{s=n+r}^{N} \eta^{(2s-1) \alpha_{s}}}{\prod_{s=n}^{n+r-1} \eta^{(2s-2) \alpha_{s} } },\\
\end{aligned}
\end{equation}
the determinant of $M_r$ is given by
\begin{equation}
\text{det}\,M_r = \prod_{s=1}^{r} \eta^{-(4n+4s-7) \alpha_{s} } >0.
\end{equation}
As a result, $\hat{\eta} B(\eta)$ is a positive definite matrix via Sylvester's criterion, with $\lambda_{k} > 0$ for all $1 \le k \le W$. Consequently, all the roots of $\text{det}\,\mathcal{Q}(z,\eta)$ are negative. The simplicity of the eigenvalues $\left\{ \lambda_{k} \right\}$ follows from the $k$-th compound matrices $C_{k}( \tilde{B})$ of $\tilde{B}(\eta)= \Gamma \hat{\eta} B(\eta) \Gamma$ with $\Gamma = \text{diag} \left( 1,-1,...,(-1)^{W-1}\right)$ that are strictly positive. The spectral radius of $C_{k}(\tilde{B})$ is given by $\rho\left( C_{k} \right) = \prod_{a=1}^{k} \lambda_{a}$ and this eigenvalue is simple from the Perron–Frobenius theorem. Thus, all $\left\{ \lambda_{k} \right\}$ are simple.

Second, we study the roots of the polynomials defined in Eqs.~(\ref{eq:p_k-recurernce-eta}) and (\ref{eq:q_k-recurernce-eta}), specifically the gap closing points. As we showed earlier, the roots of $p_{k}^{(n)}(z,\eta)$ are always negative; hence, the gap closing is only possible at $z=-1$. Due to the symmetries of $\mathcal{Q}(z,\eta)$, the polynomials $p_{k}^{(n)}(z,\eta)$ and $q_{k,n}^{(ij)}(z,\eta)$ satisfy
\begin{equation}
\label{eq:p-q-symmetries}
\begin{aligned}
p_{k}^{(n)}(z,\eta)&=z^{k}p_{k}^{(n)}(z^{-1},\eta^{-1}),\\
q_{k,n}^{(ij)}(z,\eta)&=z^{k}q_{k,n}^{(ji)}(z^{-1},\eta^{-1}).
\end{aligned}
\end{equation}
From Eq.~(\ref{eq:p_k-recurernce-eta}) and the weighted sum over $i,j$ of Eq.~(\ref{eq:q_k-recurernce-eta}), we obtain the following equality:
\begin{equation}
\label{eq:p-rec-eta-1}
p_{k+1}^{(n)}(-1,\eta) 
= d_k(\eta) p_{k}^{(n)}(-1,\eta) - s_{k-1}(\eta),\\
\end{equation}
\begin{equation}
\label{eq:p-rec-eta-2}
\text{sign} \left[s_{k} (\eta) \right] = \text{sign} \left[ p_{k}^{(n)}(-1,\eta)  \right],\\
\end{equation}
where we have introduced the following notations:
\begin{equation}
d_k(\eta) = p_{1}^{(n+k)}(-1,\eta^{\alpha_{k}}),\\
\end{equation}
\begin{equation}
s_k(\eta) = \sum_{i,j=1}^{n+k} q^{(ij)}_{k,n} (-1,\eta) \eta^{2\alpha_{k+1}(j-i)}.\\
\end{equation}
Equation~(\ref{eq:p-rec-eta-2}) demonstrated that the weighted sum of the expression in curly parenthesis in Eq.~(\ref{eq:q_k-recurernce-eta}) has a fixed sign for all $\eta > 0$ at $z=-1$. In addition, we obtain that
\begin{equation}
p_{k}^{(n)}(-1,\eta)=(-1)^{k}p_{k}^{(n)}(-1,\eta^{-1});\\
\end{equation}
thus, for odd $k$, the band gap closes if and only if $\eta =1$, since polynomials $p_{2k-1}^{(n)}(z,\eta)$ allow the following factorization due to Eq.~(\ref{eq:p-rec-eta-2}) at $z=-1$:
\begin{equation}
p_{2k-1}^{(n)}(-1,\eta)=\left( \eta - \eta^{-1}\right) P_{k}\left( \left(\eta -\eta^{-1}\right)^2\right),\\
\end{equation}
with $P_{k}(y)$ being a polynomial with a fixed sign for all $y>0$.
Since for $\eta <1$ one can use Eq.~(\ref{eq:p-q-symmetries}), we would consider $\eta >1$. Asymptotically, for $\eta \rightarrow \infty$,
\begin{equation}
 p_{2k-1}^{(n)}(-1,\eta) \underset{\eta \to \infty}{\sim}  (-1)^{k-1} \eta^{(2k-1)(2n+2k-3)}.
\end{equation}
Therefore, we obtain that
\begin{equation}
\label{eq:p-sign-odd}
\begin{aligned}
\text{sign}\left[ p_{2k-1}^{(n)}(-1,\eta)\right] &= (-1)^{k-1} ,\\
\text{sign} \left[ p_{1}^{(n+k)}(-1,\eta^{\alpha_{k}}) \right]&= \left( -1 \right)^{k},
\end{aligned}
\end{equation}
for $\eta > 1$. Since  $s_0 (\eta)>0$ and $p_2^{(n)}(-1,\eta)<0$ for all $\eta>0$, then automatically from Eqs.~(\ref{eq:p-rec-eta-1}), (\ref{eq:p-rec-eta-2}), and (\ref{eq:p-sign-odd}) we obtain that
\begin{equation}
\label{eq:p-even-ineq}
p_{2k}^{(n)}(-1,\eta) s_{2k}(\eta)>0, \,\,\,\,\forall \eta > 1.\\
\end{equation}
At $\eta=1$, Eq.~(\ref{eq:p-rec-eta-1}) simplifies as
\begin{equation}
p_{k+1}^{(n)}(-1,1) = - s_{k-1};
\end{equation}
thus, combining it with Eqs.~(\ref{eq:p-even-ineq}) and (\ref{eq:p-q-symmetries}), we obtain that $p_{2k}^{(n)}(-1,\eta) \neq 0 $ for all $\eta >0$.

Therefore for even $W$, $z=-1$ is not a solution for all $\eta >0$ and the systems with $\eta=1$ and $\eta \neq 1$ are adiabatically connected. Consequently, the topological invariant of these systems are the same and the atomic limit, considered in Sec.~\ref{sec:zak}, is valid.

\bibliography{ref}

\end{document}